\pdfoutput=1
\documentclass[12pt,preprint]{aastex}

\usepackage{hyperref}

\slugcomment{vsn 3.0, 04/07/2014}

\shorttitle{Automated Classification of WISE Periodic Variable Stars}
\shortauthors{Masci et al.}

\begin{document}


\title{Automated Classification of Periodic Variable Stars detected\\
       by the Wide-field Infrared Survey Explorer}


\author{Frank J. Masci\altaffilmark{1},
        Douglas I. Hoffman\altaffilmark{2},
        Carl J. Grillmair\altaffilmark{1},
        Roc M. Cutri\altaffilmark{1}}
%\affil{}
\email{fmasci@ipac.caltech.edu}


\altaffiltext{1}{Infrared Processing and Analysis Center, Caltech 100-22,
                 Pasadena CA 91125, USA}
\altaffiltext{2}{NASA Ames Research Center, Moffett Field, CA, 94035}


\begin{abstract}
We describe a methodology to classify periodic variable stars identified
using photometric time-series measurements constructed from the
Wide-field Infrared Survey Explorer (WISE) {\it full-mission}
single-exposure Source Databases.
This will assist in the future construction of a WISE Variable
Source Database that assigns variables to
specific science classes as constrained by the WISE observing cadence
with statistically meaningful classification probabilities.
We have analyzed the WISE light curves of 8273 variable stars
identified in previous optical variability surveys (MACHO, GCVS, and ASAS)
and show that Fourier decomposition techniques can be extended into the
mid-IR to assist with their classification.
Combined with other periodic light-curve features, this sample
is then used to train a machine-learned 
classifier based on the {\it random forest} (RF) method.
Consistent with previous classification studies of variable stars in general, 
the RF machine-learned classifier is superior to other methods in terms 
of accuracy, robustness against outliers, and relative immunity
to features that carry little or redundant class information.
For the three most common classes identified by WISE:
Algols, RR Lyrae, and W Ursae Majoris type variables,
we obtain classification efficiencies of 80.7\%, 82.7\%, and 84.5\%
respectively using cross-validation analyses, with 95\% confidence
intervals of approximately $\pm$2\%.
These accuracies are achieved at purity (or reliability) levels of
88.5\%, 96.2\%, and 87.8\% respectively, similar to that achieved
in previous automated classification studies of periodic variable stars.
\end{abstract}





\section{Introduction}

The Wide-field Infrared Survey Explorer \citep[WISE;][]{wright10}
mapped the entire sky in four bands centered at wavelengths of 3.4,
4.6, 12, and 22 $\mu$m (hereafter W1, W2, W3, and W4) from
January 7, 2010 to February 1, 2011 that spanned both cryogenic and
post-cryogenic phases.
WISE conducted its survey from a sun-synchronous polar orbit
using a 40 cm cryogenically cooled telescope equipped with
four 1024 $\times$ 1024 infrared (IR) array detectors that
simultaneously imaged the same 47\arcmin~$\times$ 47\arcmin~field-of-view
in all bands. The WISE survey strategy alternated stepping the scan
path forward and backward on subsequent orbits in an asymmetric pattern that
approximately matched the $\approx1\arcdeg$ per day orbital precession rate.
In this way, each point near the {\it ecliptic plane} was observed on every
other orbit, or every 191 minutes, yielding typically 12 independent exposures
over one day. The number of exposures increases with ecliptic latitude,
reaching over 6000 at the ecliptic poles.

WISE surveyed the sky approximately 1.5 times during its cryogenic phase
that ended on September 29, 2010. Data continued to be collected 
for another four months to support the discovery of Near Earth Objects
\citep[the NEOWISE program;][]{Mainzer11}. During this post-cryogenic phase,
70\% of the sky was scanned, with only the W1 and W2 detectors returning
scientifically useful data. Overall, WISE covered the full sky slightly more
than twice, with each sky-coverage separated by approximately six months. 
When all the mission data are combined, the median single-exposure
depth-of-coverage on the ecliptic becomes $\sim 24$ and the effective
observation timespan (without the six month phase shift) $\sim2$ days.

The WISE Single-Exposure Source Databases \citep{cutri12} contain the
photometry from each individual WISE exposure. These
offer a unique opportunity to search for variable stars over the
entire sky at wavelengths that are relatively immune to dust extinction.
In particular, the most common pulsational variables
(e.g., Cepheids, RR Lyrae, and Miras) have served as standard candles
that enabled measurements of the size scale of the Milky Way
and the Universe \citep[e.g.,][]{tammann08}. Products from
previous WISE data releases
have already had a significant impact on the calibration of the RR Lyrae
period-luminosity relation at mid-IR wavelengths
\citep{klein11,madore13,dambis14,klein14}.
An all-sky census of pulsating variables offers an opportunity to improve our
understanding of Milky Way tomography and the distribution of dark
matter in the Galaxy through their association with relic streams from
disrupted star clusters and dwarf galaxies \citep{grillmair10}.
Pulsational variables are also crucial for understanding stellar birth,
structure, mass loss, and evolution \citep{eyer08}.
On the other hand, variability associated with eclipsing binaries
(e.g., Algol, $\beta$ Lyrae, and W Ursae Majoris types) provide
laboratories for probing accretion, mass transfer, binary evolution,
and exoplanets \citep[for a review, see][]{percy07}.

The deluge of data from current and future time-domain surveys presents
an enormous challenge for human-based vetting, classification,
and follow-up. Fortunately, computers and efficient machine-learning (ML)
algorithms are starting to revolutionize the taxonomic problem.
A broad class of ML methods are generically referred to as ``supervised''.
These methods entail defining a set of rules or models that best
describe the relationship between the properties (features) of a
data set and some known outcomes (e.g., classifications),
and then using this ``trained'' mapping to predict the outcomes for new data.
On the other hand, ``unsupervised'' methods attempt to blindly identify
patterns and structures amongst the features of a data set, that is, with no
pre-labelled classes or outcomes to learn from.
For an overview of ML methods in general, see \citet{hastie09}.
For a review of ML applications in astronomy, see \citet{ball10}.
ML also provides a statistical framework to make
probabilistic statements about
the class(es) that a particular object with a set of
observables or features could belong to. Given the same training model,
these statements are also
reproducible and deterministic, whereas a human-based classification
approach is not.

Supervised ML techniques have gained considerable
popularity in the automated classification of variable stars
\citep{wozniak04,eyer05,deboss07,mahabal08,blomme10,richards11}.
In particular, \citet{richards11} compared
several classification frameworks for variable stars selected from multiple
surveys and concluded that the {\it random forest} ML classifier 
\citep[a tree-based technique popularized by][]{breiman01}
generally performed best.
Due to their flexibility and robustness, {\it random forests} have also
attained popularity in the real-time discovery and classification of
astronomical transients in general \citep[e.g.,][]{bloom12,morgan12,brink13}.
Guided by its success, we adopt the {\it random forest} method as 
the primary ML classifier in this study.

The WISE All-Sky Release Source Catalog \citep{cutri12} contains
approximately 8 million sources where likely flux variables were flagged
using a methodology similar to that described in \cite{hoffman12}.
We are in the process of constructing the WISE Variable Source Database 
(WVSDB) that builds upon this basic variability flagging in the Catalog.
This database will contain confirmed variable sources with their light curves,
derived properties such as periods and amplitudes, and where appropriate,
the probabilities of belonging to specific known variability classes.

The first step towards the construction of the WVSDB is
a framework that can automatically and probabilistically 
classify variables using features and diagnostics derived from the 
WISE single-exposure time-series measurements.
This paper describes the methodology behind this framework, starting
with the construction of a training (or ``truth'') sample
leveraged on known variables classified in
previous optical surveys and cross-matched to WISE to
define robust mid-IR light curve features for classification.
In particular, we show how Fourier decomposition techniques can be extended
into the mid-IR to define the relevant features for discriminating between
the various classes. This training sample is then used to fit and validate
the popular {\it random forest} machine-learning technique to
assist with the future classification of WISE flux variables for the WVSDB.

This paper is organized as follows. In Section~\ref{cscheme} we define the
variability classes that WISE is most sensitive to.
Section~\ref{tsampf} describes
the construction of the training sample and selection of the mid-IR
light curve features. An analysis of these features for the various
classes of interest is presented in Section~\ref{fan}. The {\it random forest}
machine-learning method for automated classification is described and
evaluated in Section~\ref{ml}, where we also compare this method to other 
state-of-the-art machine-learning methods.
Section~\ref{cpred} gives an overview of the WVSDB, the classification plan,
and how a feedback mechanism based on ``active-learning'' could be
used to allow for selection biases in the input training sample.
Results are summarized in Section~\ref{conc}.

\section{Classification Scheme}\label{cscheme}

The WISE {\it full mission} baseline and observing cadence is well-suited for
studying periodic variable stars with periods of $\lesssim 3$ days,
where 3 days is approximately the maximum period that can be recovered
using our period estimation technique (Section ~\ref{fgen}) for light curves
constructed from observations within a few tens of degrees of the ecliptic.
The most common variables in this category
are RR Lyrae (RR Lyr) pulsational variables and 
Algol, $\beta$ Lyrae, and W Ursae Majoris (W UMa)
eclipsing binaries.
Optical surveys generally find W UMa variables (a class of contact binaries)
to be the most abundant,
comprising $\sim$95\% of all variable stars in the solar neighborhood
\citep{eggen67,lucy68}. Despite their small variability amplitudes and
relatively weak emission in the mid-infrared, WISE is sensitive to 
the brightest W UMa variables at the high end of the amplitude distribution.
$\beta$ Lyrae eclipsing binaries are a class of semi-detached binary stars
where one member of the pair fills the Roche lobe of the other.
Previous optical studies have
shown that $\beta$ Lyrae are generally difficult to separate from
Algol-type (detached) eclipsing binaries based on light-curve shape alone
\citep{malkov07,hoffman08}. Our analyses of the WISE light curve features
(Section~\ref{fan}) also show these variables to largely overlap
with Algols, and also perhaps with W Uma variables to some extent.
The degeneracies can only be resolved with supplementary spectral information. 
Therefore, we are left with three broad periodic variable star
classes that are best suited for WISE's observing constraints:
Algols (with the inclusion of many $\beta$ Lyrae); W UMa; and RR Lyr. 

Periodic variables that are {\it not} assigned to any of these three 
broad classes (according to some probability threshold; see
Section~\ref{cpred}) will be initially flagged as ``unknown'';
for example, Cepheid variables.
They may be reclassified and associated with new classes
(not in the initial training sample described in Section~\ref{tsamp}) 
if more objects with similar features are identified following a first
classification pass. Subsequent classification passes will use refined
training samples augmented with new classes using an
``active-learning'' approach. Details are given in Section~\ref{al}.

\section{Training Sample and Feature Selection}\label{tsampf}

\subsection{Training Sample Definition}\label{tsamp}

In order to calibrate and validate a classification method for WISE
light curves, we assembled a ``truth'' list of variable stars that
were previously classified with measured periods
from a number of optical variability surveys.
This list includes all eclipsing
binaries and RR Lyr stars in the General Catalogue of Variable Stars
\citep[GCVS;][]{samus13}; the MACHO Variable Star Database
\citep{alcock03}, and the All-Sky Automated Survey
\citep[ASAS;][]{pojmanski06}. This list of known variables was then
positionally cross-matched to the WISE All-Sky Source Catalog \citep{cutri12}.
A match-radius tolerance of $2.5\arcsec$ was used and the brightest
WISE source was selected if multiple matches were present.
Furthermore, only sources with a WISE Catalog variability flag 
\citep[][section IV.4.c.iii]{cutri12} of
$var\_flg\geq 6$ in the W1 band were retained. This criterion ensured that
the WISE source had a relatively good time-averaged photometric
signal-to-noise ratio (with typically S/N $\gtrsim 6$ in the W1
single-exposures) and a high likelihood of also being variable in the mid-IR.
After cross-matching, 1320 objects were rejected because they were labeled
as either ambiguous (assigned to two or more classes), uncertain, or
had different classifications between the optical catalogs. A further
2234 objects had duplicate entries amongst the optical catalogs (assigned
to the same class) and one member of each pair was retained. In the end, we
are left with a training sample of 8273 known variables with WISE
photometry. Of these 8273 objects, 1736 are RR Lyr, 3598 are Algol-type
eclipsing binaries, and 2939 are W UMa-type eclipsing binaries according
to classifications reported in previous optical variability surveys.

We assume here that the bulk of classifications reported in the
ASAS, GCVS, and MACHO surveys (primarily those objects labeled as
Algol, RR Lyr, or W Uma) are correct, or accurate enough as to not
significantly affect the overall core-class definitions in the WISE
parameter (feature) space. The removal of objects with ambiguous and
uncertain classifications, or with discrepant classifications between
the catalogs is expected to have eliminated the bulk of this uncertainty.
A comparison of our classification performance to other studies
that also used classifiers trained on similar optical catalogs
(Section~\ref{cv}) shows that possible errors in the classification labels
of the training sample are not a major source of uncertainty.
However, these errors cannot be ignored.
  
There is no guarantee that the training sample described here
represents an unbiased sample of all the variable types that WISE can recover
or discover down to fainter flux levels and lower S/N ratios over the
entire sky. This training sample will be used to construct an initial
classifier to first assess general classification performance
(Section~\ref{cv}). In future, this classifier will be used to
initially classify WISE flux-variables into the three broad variability
classes defined above. Inevitably, many objects will remain
unclassified in this first pass. To mitigate training sample biases and
allow more of the WISE-feature space to mapped and characterized,
we will use an active-learning framework to iteratively refine the
training sample as objects are accumulated and (re)classified.
This will allow more classes to be defined as well as sharpen
the boundaries of existing ones, hence alleviating
possible errors in the input classification labels (see above).
Details are discussed under our general classification plan
in Section~\ref{cpred}.

\subsection{Mid-IR Light Curve Feature Generation}\label{fgen}

A requirement of any classification system is a link between 
the features used as input and the classes defined from them.
We review here the seven mid-IR light curve features that we found work
best at discriminating between the three broad classes that
WISE is most sensitive to (defined in Section~\ref{cscheme}).
These were motivated by previous variable star classification
studies \citep[e.g.,][]{kinemuchi06,deb09,richards11}

We first extracted the time-resolved mid-IR photometry and constructed
light curves for all sources in our
training sample from the WISE All-Sky, 3-Band, and Post-Cryo
Single-Exposure Source
Databases\footnote{\texttt{\url{http://irsa.ipac.caltech.edu/Missions/wise.html}}}.
Light curves were constructed using only the W1 band.
This is because W1 generally has better sensitivity than W2 for the
variable stars under consideration. For each source, a period was estimated
using the generalized Lomb-Scargle periodogram
\citep[][see Section~\ref{fan} for this choice]{zechmeister09,scargle82}.
The light curves were phased to the highest peak in the periodogram
that fell in the period range: 0.126 day ($\sim$ 3 hours) to 10 days.
The lower value corresponds to the characteristic WISE single-exposure
sampling and the upper limit is based on the known periods of Algol-type
binaries that could be recovered by WISE given the typical time-span of
observations, with a generous margin. These recovered periods ($P_r$)
constitute our first feature to assist with classification.
Section~\ref{fan} compares these estimates to the available periods
derived from optical light curves.

The second feature derived from the WISE light curves in our training
sample is the Stetson-$L$ variability index \citep{stetson96,kim11}.
This index quantifies the degree of synchronous variability between two bands.
Because the band W1 and W2 measurements
generally have the highest S/N ratio for objects in the classes
of interest, only these bands are used in the calculation of this index.
The Stetson-$L$ index is the scaled product of the
Stetson $J$ and $K$ indices:
\begin{equation}\label{stetL}
  L = \frac{JK}{0.798}.
\end{equation}
Stetson-$J$ is a measure of the correlation between two bands ($p$, $q$;
or W1, W2 respectively) and is defined as
\begin{eqnarray}
  J & = & \frac{1}{N} \sum_{i=1}^N \mbox{sgn}(P_i) \sqrt{|{P_i}|}, \\
  P_i & = & \delta_{p}(i)\delta_{q}(i), \\
  \delta_p(i) & = & \sqrt{\frac{N}{N-1}} \frac{m_{p,i} -
                    \bar{m}_{p}}{\sigma_{p,i}}, \\
  \bar{m}_{p} & = & \frac{1}{N}\sum_{i=1}^N m_{p,i},                 
\end{eqnarray}
where $i$ is the index for each data point, $N$ is the total
number of points, sgn($P_i$) is the sign of $P_i$,
$m_{p,i}$ is the photometric magnitude of flux measurement $i$ in band
$p$, and $\sigma_{p,i}$ is its uncertainty.
The Stetson-$K$ index is a measure of the kurtosis of the
magnitude distribution and is calculated
by collapsing a single-band light curve:
\begin{equation}
K = \frac{1}{\sqrt{N}} \frac{\sum_{i=1}^{N}
    |\delta(i)|}{\sqrt{\sum_{i=1}^{N} \delta(i)^2}}.
\end{equation}
For a pure sinusoid, $K\simeq0.9$, while for a Gaussian magnitude
distribution, $K\simeq0.798$. This is also the scaling
factor in the $L$-index (equation~\ref{stetL}).

Our third derived feature is the magnitude ratio
\citep[$MR$;][]{kinemuchi06}. This measures the fraction
of time a variable star spends above or below its median
magnitude and is useful for distinguishing between variability from
eclipsing binaries and pulsating variables.  This is computed using the
magnitude measurements $m_i$ for an individual band and is defined as
\begin{equation}\label{mr}
MR = \frac{\mbox{max}(m_i) - \mbox{median}(m_i)}{\mbox{max}(m_i) -
     \mbox{min}(m_i)}.
\end{equation}
For example, if a variable star spends $>$50\% of its time at
near constant flux that only falls occasionally,
$MR\approx 1$. If its flux rises occasionally, $MR\approx 0$.
A star whose flux is more sinusoidal will have $MR\approx 0.5$.

The remaining four features are derived from a Fourier decomposition of the
W1 light curves. Fourier decomposition
has been shown to be a powerful tool for variable star classification
\citep{deb09,rucinski93}. To reduce the impact of noise and outliers
in the photometric measurements, we first smooth a mid-IR light curve
using a local non-parametric regression fit with Gaussian kernel of
bandwidth ($\sigma$) 0.05 days. We then fit a 5$th$ order Fourier series to the
smoothed light curve $m(t)$, parameterized as
\begin{equation}\label{eq:fourier}
m(t) = A_0 + \sum_{j=1}^5 A_j \cos[2\pi j \Phi(t) + \phi_j]
\end{equation}
where $\Phi(t)$ is the orbital phase at observation
time $t$ relative to some reference time $t_0$ and is
computed using our recovered period $P_r$ (see above) as
\begin{equation}
\Phi(t) = \frac{t - t_0}{P_r} - \mbox{int}\left(\frac{t-t_0}{P_r}\right),
\end{equation}
where ``int'' denotes the integer part of the quantity and
$0\leq\Phi(t)\leq1$. The parameters that are fit in equation
(\ref{eq:fourier}) are the amplitudes $A_j$ and phases $\phi_j$.
The quantities that we found useful for classification
\citep[see section~\ref{fan} with some guidance from][]{deb09}
are the two relative phases
\begin{equation}\label{eq:phi1}
\phi_{21} = \phi_2 - 2\phi_1
\end{equation}
\begin{equation}\label{eq:phi2}
\phi_{31} = \phi_3 - 3\phi_1,
\end{equation}
and the absolute values of two Fourier amplitudes: $|A_2|$ and $|A_4|$.

To summarize, we have a feature-vector consisting of seven metrics
for each mid-IR light curve in our training sample: the recovered
period $P_r$, the magnitude ratio $MR$ (equation~\ref{mr}), the
Stetson-$L$ index (equation~\ref{stetL}), and the four Fourier
parameters: $|A_2|$, $|A_4|$, $\phi_{21}$, and $\phi_{31}$.
Together with available class information from the literature
(the dependent variable), we constructed a data matrix consisting
of 8273 labeled points (``truth'' samples) in a seven-dimensional space.
Section~\ref{ml} describes how this data matrix is used
to train and validate a machine-learned classifier.

\section{Preliminary Analysis of Features for Classification}\label{fan}

Before using our training sample to construct an automated classification
algorithm, we present here a qualitative analysis of the derived light-curve
features across the three different classes of interest. That is, how well
they perform individually and in combination, in a qualitative sense,
for discriminating between classes. The relative feature importance
across (and within) classes will be explored in more detail using metrics
generated during the classifier-training phase in Section~\ref{import}.

The accuracy of period recovery is an important factor in the classification
process, in particular since this metric is also used (indirectly) to
derive the features from Fourier decomposition.
Given all three target classes overlap at least partially in period,
it is important to minimize any period aliasing as much as possible,
that is, any inadvertent phasing of the time-series data to an integer
harmonic of the true period.
The generalized Lomb-Scargle periodogram \citep[GLS;][]{zechmeister09} was
superior to other methods in recovering the correct period and minimizing
period aliasing. The other methods we explored were the standard
Lomb-Scargle algorithm; phase dispersion minimization method
\citep[PDM;][]{stellingwerf78};
multiharmonic analysis of variance \citep[AOV;][]{c-z98}; and the
string-length algorithm \citep{lafler65}. The GLS method recovered
the correct period for the largest number of variables and
minimized period aliasing (for mostly the RR Lyr class; see below).
This is likely due to the relatively
sparse temporal sampling of many WISE sources where GLS is most
robust. GLS also has the advantage of incorporating measurement
uncertainties in the calculation, whereas many other methods do not.

%\placefigure{fig:period}

\begin{figure}[hbtp]
\begin{center}
\includegraphics[scale=0.8]{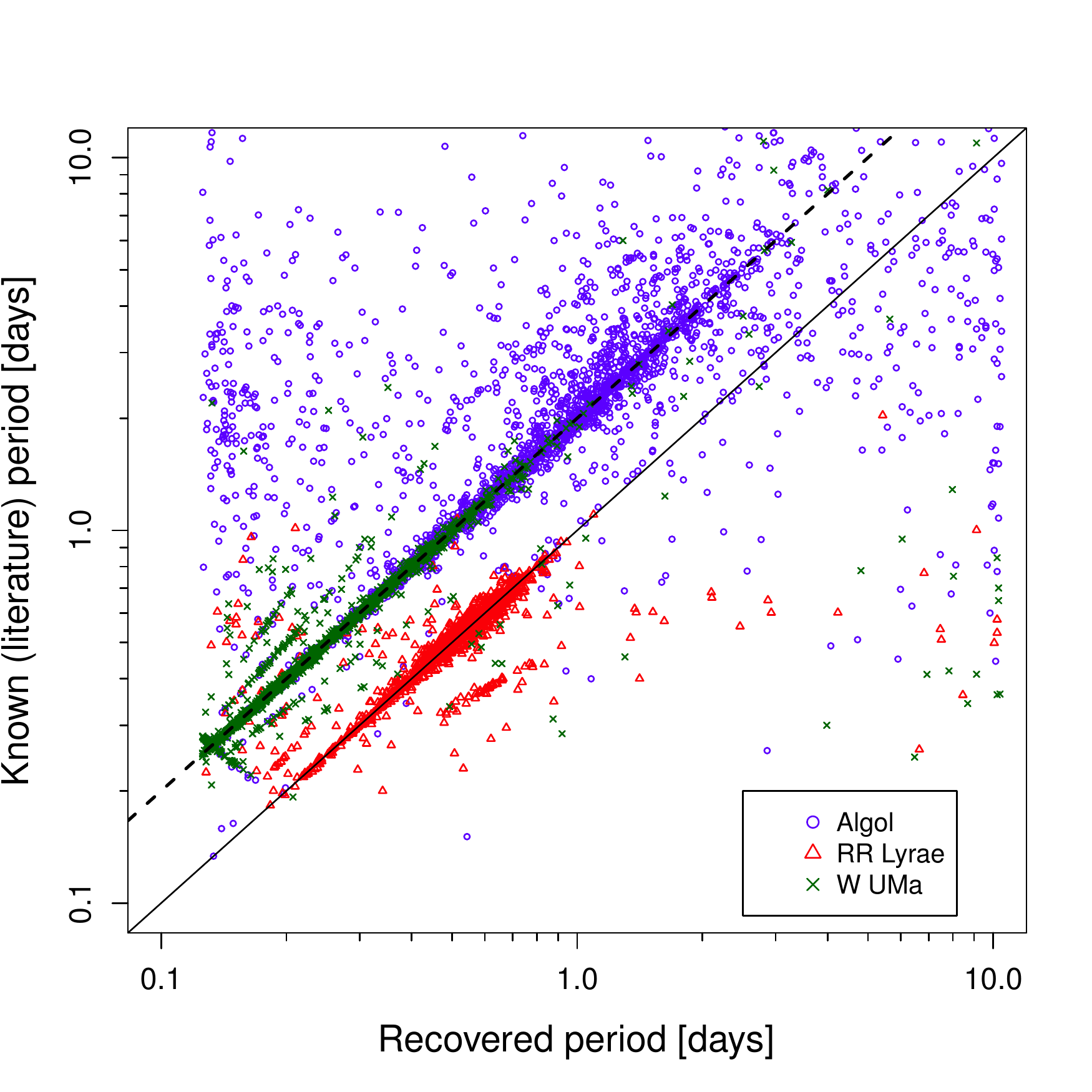}
\caption{The recovered periods using the W1 light curves compared to
known (prior) periods from previous optical variability surveys for
the three broad classes in our training sample.
The Algol class ({\it blue} circles) consists primarily of detached eclipsing
binaries but also includes some semi-detached systems. The W Uma variables
({\it green} crosses) are primarily contact binaries,
and the RR Lyr ({\it red} triangles)
are the only pulsational variables considered in this study. The solid
line is the line of equality and the dashed line is where the recovered
period is half the known period -- the half-period aliasing effect discussed
in Section~\ref{fan}.}
\label{fig:period}
\end{center}
\end{figure}

As shown in Figure \ref{fig:period}, period recovery is good for periods
of less than $\sim 2.5$ days. Longer periods are more difficult to recover
from WISE data due to the observing cadence, as is evident by the
increased scatter at longer periods.
Nearly all the RR Lyr variables are recovered at the fundamental
period, while the Algol and W UMa variables
are recovered at half the period. This separation
arises from the fact that eclipsing systems usually have two minima per cycle
(the primary and seconday eclipses)
while pulsating variable stars have only one.
This half-period aliasing for the periods of eclipsing binaries
does not impact their classification (or separability from
pulsating variables in general) since we find they can
be reliably distinguished using other features
(see below). Thus, once an eclipsing binary has been identified
using all the available light curve features, their
measured period can be doubled to recover the correct period.
One should also note several alias
groupings in Figure \ref{fig:period}, particularly for the W UMa class.
These are due to the sinusoidal nature of their light curves
and their relatively short periods ($\lesssim0.4$ day) where
sparse-sampling of the WISE data can significantly affect period recovery.

\begin{figure}[hbtp]
\begin{center}
\includegraphics[scale=0.9]{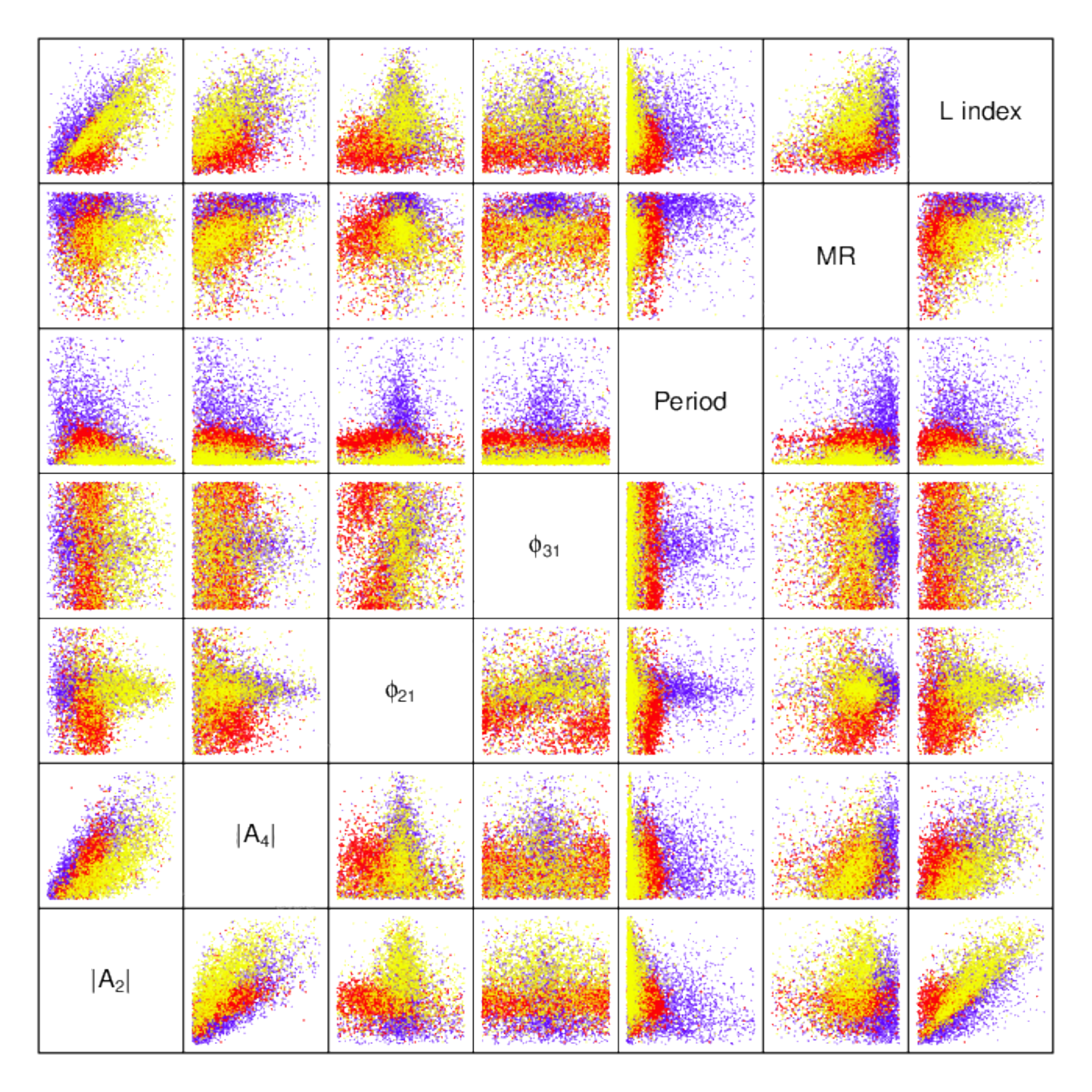}
\caption{Matrix of scatter plots for all possible pairs of metrics in
our seven parameter feature vector. Labels for the {\it x} and
{\it y} axes (or columns and rows respectively) are indicated
along the diagonal. {\it Blue} points are Algol variables (detached eclipsing
binaries but also including semi-detached systems); {\it red} points are
RR Lyrae; and {\it yellow} points are W Uma variables (contact binaries).}
\label{fig:featureplot}
\end{center}
\end{figure}

Each feature in our seven-dimensional feature vector is compared against
every other in Figure \ref{fig:featureplot} for the three classes of
interest in our training sample. There are 21 unique sets of feature pairs.
This scatter-plot matrix provides both a qualitative
sense of the degree of correlation between features as
well as class separability in each two-dimensional
projection. Feature correlations and possible redundancies are
further explored in Section~\ref{fcor}.
The features that tend to separate the classes relatively well
involve combinations of the Fourier amplitudes 
($|A_2|$, $|A_4|$) and relative phase parameters
($\phi_{21}$, $\phi_{31}$), but the separation is also relatively
strong in the $L$-index versus magnitude ratio ($MR$) plane.  
We expand on the details for three pairs of features below.

%\placefigure{fig:featureplot}

\begin{figure}[hbtp]
\begin{center}
\includegraphics[scale=0.8]{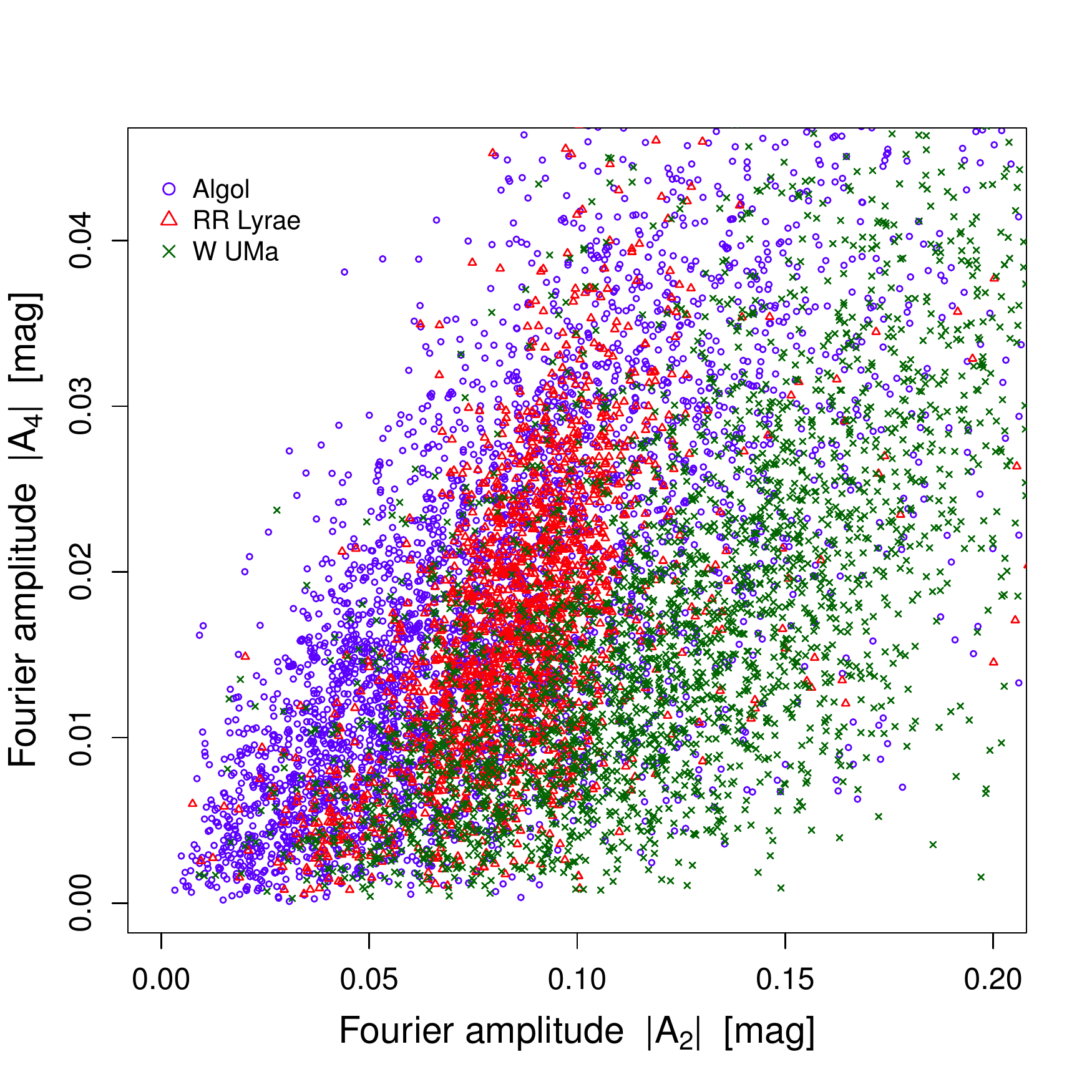}
\caption{The absolute value of the second and fourth
Fourier amplitudes from fitting equation (\ref{eq:fourier}) to the band W1
light curves for the three classes in our training sample.}
\label{fig:A}
\end{center}
\end{figure}

\begin{figure}[hbtp]
\begin{center}
\includegraphics[scale=0.8]{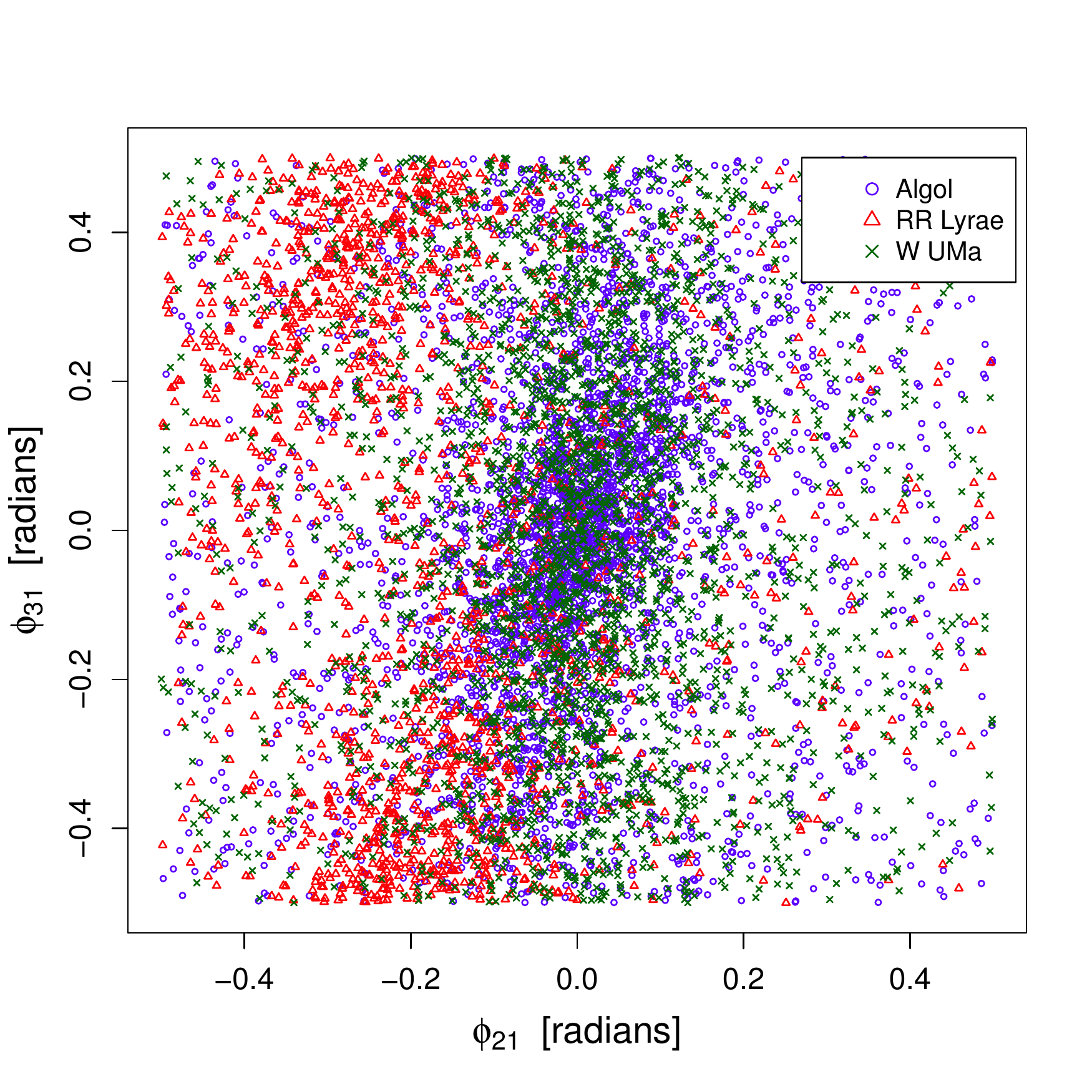}
\caption{The relative phase parameters (equations \ref{eq:phi1} and
\ref{eq:phi2}) from fitting equation (\ref{eq:fourier}) to the band
W1 light curves for the three classes in our training sample.}
\label{fig:phi}
\end{center}
\end{figure}

\begin{figure}[hbtp]
\begin{center}
\includegraphics[scale=0.8]{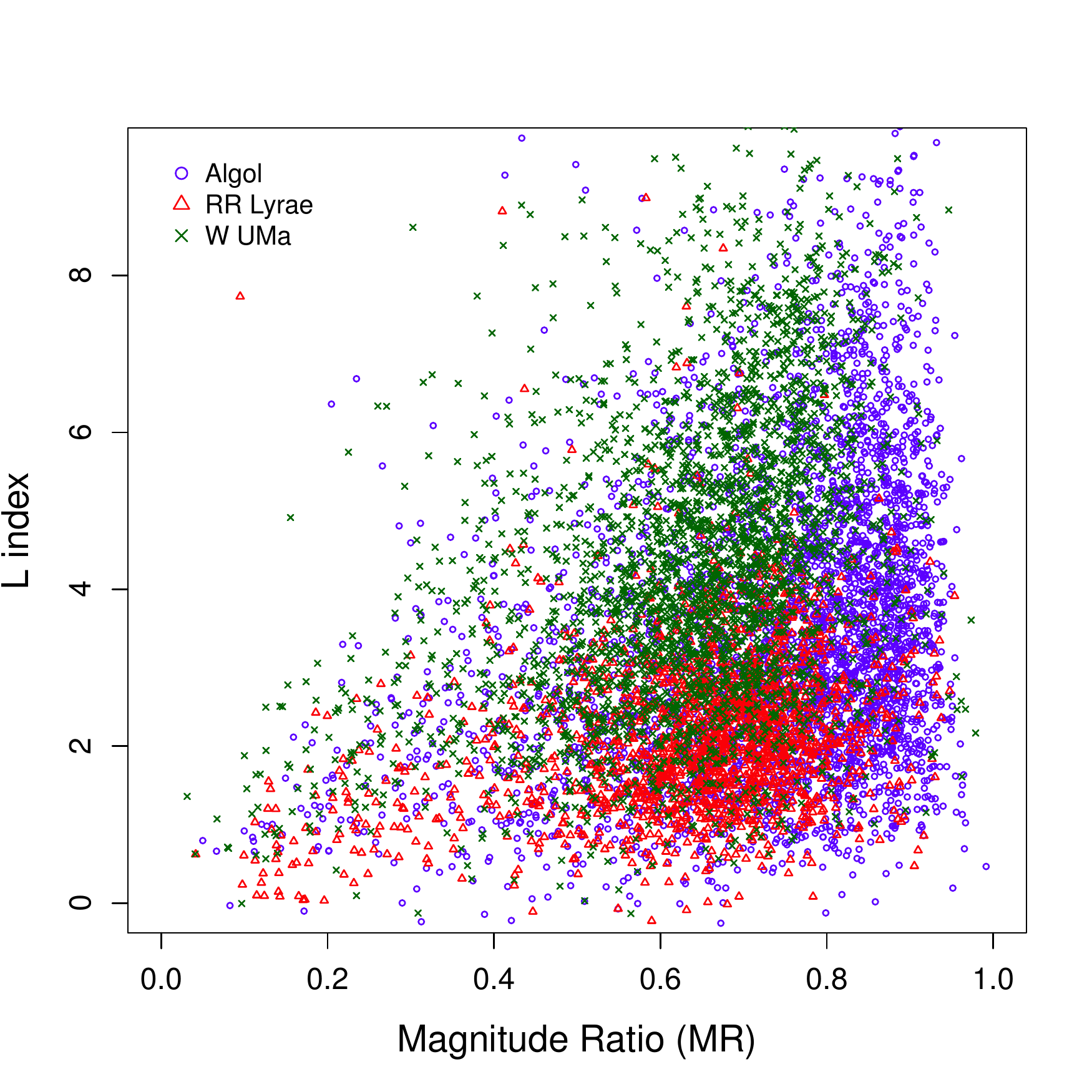}
\caption{The Stetson $L$-index (equation~\ref{stetL}) versus the
Magnitude Ratio (equation~\ref{mr}) for the
three classes in our training sample.}
\label{fig:MR-L}
\end{center}
\end{figure}

Figures \ref{fig:A} and \ref{fig:phi} show the distribution of the
Fourier amplitudes and relative phase parameters for each class.
The RR Lyr and W UMa classes in particular appear to partially overlap
in each 2-D plane formed by each pair of parameters. This is
because many RR Lyr, especially those of the
RRc subclass, have nearly sinusoidal light curves that are very similar
to some W UMa variables. The periods of these two classes
(Figure \ref{fig:period}) also overlap to some extent.
Figure \ref{fig:MR-L} shows the benefit of including the magnitude ratio and
Stetson $L$-index to assist in distinguishing RR Lyr from the
eclipsing binary (Algol and W UMa) classes in general,
which may not be achieved using the Fourier amplitudes
or phase parameters alone.

%\placefigure{fig:A}

%\placefigure{fig:phi}

%\placefigure{fig:MR-L}

The Algol class appears to isolate itself rather well from the RR Lyr 
and W Uma classes in Figures \ref{fig:A} and \ref{fig:MR-L}, and only from
the RR Lyr class in Figure \ref{fig:phi}. This is due to the asymmetrical
nature of the Algol-type light curves and the fact that their primary
and secondary minima are separated by an
orbital phase of 0.5. This is common amongst Algol-type eclipsing binaries
since most of them have orbital eccentricities of approximately zero.
This also explains why there is a tight clump centered at
$\phi_{21}\approx0$ and $\phi_{31}\approx0$ in Figure~\ref{fig:phi}.
There is however some small overlap between Algols and
the RR Lyr and W UMa classes (which is smaller than that
between the RR Lyr and W Uma classes). This primarily
occurs for the shorter period Algols with
similar depths in their primary and secondary eclipses, indicating
the component stars are similar in nature. For these, the light curves 
become more sinusoidal and indistinguishable from the other classes.

From this preliminary exploratory analysis of the light curve features
(using simple pairwise comparisons), it is clear that we need to explore
class separability using the full joint seven-dimensional feature space,
with some quantitative measure for assigning class membership. This
is explored in the following sections. In particular, the relative
feature importance is revisited and explored in more detail in
Section~\ref{import}.

\section{Machine-Learned Supervised Classification Framework}\label{ml}

The classes and their features defined above form the basis of
a ``supervised'' classification framework. This method
uses a sample of objects (here variable sources) with known
classes to train or learn a non-parametric function (model) that describes
the relationship between the derived features and these classes.
This sample of labeled classes is referred to as the training sample.
Our ultimate goal is to use this model to automatically
predict the most probable class of future objects from its derived features,
or in general, the probability that it belongs to each of the pre-defined
classes. These probabilities quantify the degree to which
a particular object could belong to specific class, therefore making
the classification process less subjective, or more open to
interpretation and further analysis. Section~\ref{train} describes
how these probabilities are defined.

Many previous studies have used machine learning (ML) methods to classify
variable stars from their photometric time series, in particular, in
large surveys capable of identifying 20 or more variability classes.
The intent has been to develop a generic classifier that is accurate,
fast and robust, and can be used to classify objects from surveys
other than those used to construct the classifier.
\citet{eyer.etal08}, \citet{richards11}, and \citet{long12}
discuss some of the challenges on this front.
The main challenge here is the presence of survey-dependent systematics,
for example, varying cadence, flux sensitivity, signal-to-noise ratio,
number of epochs, etc. This heterogeneity introduces systematic
differences in the derived light-curve features, leading to biased
training models and degraded classifier performance when attempting to
classify new objects. A related issue is class-misrepresentation in the
training model, i.e., where classes are inadvertently omitted because of
limitations in the specific survey(s) used to construct the
training sample. This will impact classifications for other surveys
whose properties may allow these additional classes to be detected.
Training a classifier using a subset of labeled (pre-classified) objects
drawn from the {\it same} target sample for which bulk classifications are
required (i.e., with similar properties) minimizes these biases, but does not
eradicate them. Methods to mitigate training-sample biases for the
classification of WISE flux-variables are discussed in Section~\ref{al}.

Some of the ML methods used to classify variable stars
include support vector machines \citep{wozniak04,deboss07},
Kohonen self-organizing maps \citep{brett04},
Bayesian networks and mixture-models \citep{eyer05,mahabal08},
principle component analysis \citep{deb09},
multivariate Bayesian and Gaussian mixture models
\citep{blomme10,blomme11} for the {\it Kepler} mission,
and thick-pen transform methods \citep{park13}.
\citet{deboss07} explored a range of methods applied to
several large surveys that included {\it Hipparcos} and {\it OGLE}:
artificial neural networks, Bayesian networks, Gaussian mixture models, and
support vector machines. All these methods appear to
achieve some level of success; however, using the same input data,
\citet{richards11} and \citet{dubath11} explored the performance of
tree-based classification schemes that include {\it random forests} and
found these to be generally
superior to the methods in \citet{deboss07} in terms of accuracy,
robustness to outliers, ability to capture complex structure in the
feature space, and relative immunity to irrelevant and redundant features.

Unlike the complex heterogeneous nature of the
many-class/multi-survey classification problem noted above, the good
overall homogeneity of
the WISE survey provides us with a well-defined sample of uniformly
sampled light-curves from which we can train and tune a ML classifier
and use it for initial classification in future. As discussed in
Section~\ref{tsamp} and further in Section~\ref{al}, these initial 
classifications will be used to refine the classifer and mitigate 
possible biases in the input training sample.
Below, we focus on training and validating a {\it random forest}
(RF) classifier, then compare its performance to some other
state-of-the-art methods: artificial neural networks (NNET),
$k$-Nearest Neighbors ($k$NN), and support vector machines (SVM).
We do not delve into the details of these other methods, as our intent is
simply to provide a cross-check with the RF classifier.

\subsection{Classification using Trees and Random Forests}\label{rf}

ML methods based on classification and regression trees (CART)
were popularized by \citet{breiman84}.
Decision trees are intuitive and simple to construct.
They use recursive binary partitioning of a feature space by splitting
individual features at values (or decision thresholds) to create
disjoint rectangular regions -- the nodes in the tree. The tree-building
process selects both the feature and threshold at which
to perform a split by minimizing some measure of the inequality
in the response between the two adjacent nodes (e.g., the
fractions of objects across all {\it known} classes, irrespective of
class). The splitting
process is repeated recursively on each subregion until some terminal
node-size is reached ({\it nodesize} parameter below).

Classification trees are powerful non-parametric classifiers that
can deal with complex non-linear structures and dependencies
in the feature space. If the trees are sufficiently deep, they
generally yield a small bias with respect to the true model that
relates the feature space to the classification outcomes. Unfortunately,
single trees do rather poorly at prediction since they lead to a
high variance, e.g., as encountered when overfitting a
model to noisy data. This is a consequence of the hierarchical structure
of the tree: small differences in the top few nodes can produce
a totally different tree and hence wildly different outcomes.
Therefore, the classic bias versus variance tradeoff problem
needs to be addressed. To reduce this variance, \citet{breiman96}
introduced the concept of {\it bagging} (bootstrap aggregation).
Here, many trees ($N_{\rm tree}$ of them) are built from randomly selected
(bootstrapped) subsamples of the training set and the
results are then averaged. To improve the accuracy of the final averaged
model, the {\it random forest} (RF) method \citep{breiman01} extends
the bagging concept by injecting further randomness into the
tree building process. This additional
randomness comes from selecting a {\it random subset} of the input
features ($m_{\rm try}$ parameter below) to consider in the
splitting (decision) process at each node in an individual tree. 
This additional randomization ensures the trees are more
de-correlated prior to averaging and gives a lower variance
than bagging alone, while maintaining a small bias.
These details may become clearer in section~\ref{train} where
we outline the steps used to tune and train the RF classifier.
For a more detailed overview, we refer the interested reader to ch.15 of
\citet{hastie09} and \citet{breiman04}. 

{\it Random forests} are popular for classification-type problems and are
used in many disciplines such as bioinformatics, Earth sciences, economics,
genetics, and sociology.
They are relatively robust against
overfitting and outliers, weakly sensitive to choices of tuning parameters,
can handle a large number of features, can achieve good accuracy (or minimal
bias and variance), can capture complex structure in the feature space,
and are relatively immune to irrelevant and redundant features.
Furthermore, RFs include a mechanism to assess the relative importance of each
feature in a trivial manner. This is explored in section~\ref{import}.
For our tuning, training, and validation analyses, we use tools from the
\texttt{R} statistical software environment\footnote{\texttt{R} is 
freely-available at \texttt{\url{http://cran.r-project.org}}}.
In particular, we make extensive use
of the \texttt{caret} (Classification and Regression Training) machine
learning package \citep{kuhn08}, version 5.17-7, August 2013.

\subsection{Feature Collinearity and Redundancy Checks}\label{fcor}

As mentioned earlier, the RF method is relatively immune to features
that are correlated with any other feature or some linear combination
of them, i.e., that show some level of redundancy in ``explaining''
the overall feature space. However, it is recommended that features
that are strongly correlated with others in the feature set be removed
in the hope that a model's prediction accuracy can be ever slightly
improved by reducing the variance from possible over-fitting.
Given our training classes contain a relatively large number of
objects ($\gtrsim 1200$) and our feature set consists of only seven
predictor variables, we do not expect over-fitting to be an issue,
particularly with RFs. Our analysis of feature collinearity is
purely exploratory. Combined with our exploration of relative
feature importance in Section~\ref{import}, these analyses fall under
the general topic of ``feature engineering'' and are considered common
practice prior to training any ML classifier.

To quantify the level of redundancy or correlation amongst our $M$ features,
we used two methods: (i) computed the full pair-wise correlation
matrix and (ii) tested for general collinearity by regressing each feature
on a linear combination of the remaining $M-1$ features and examining
the magnitude and significance of the fitted coefficients.

\begin{figure}[hbtp]
\begin{center}
\includegraphics[scale=0.9]{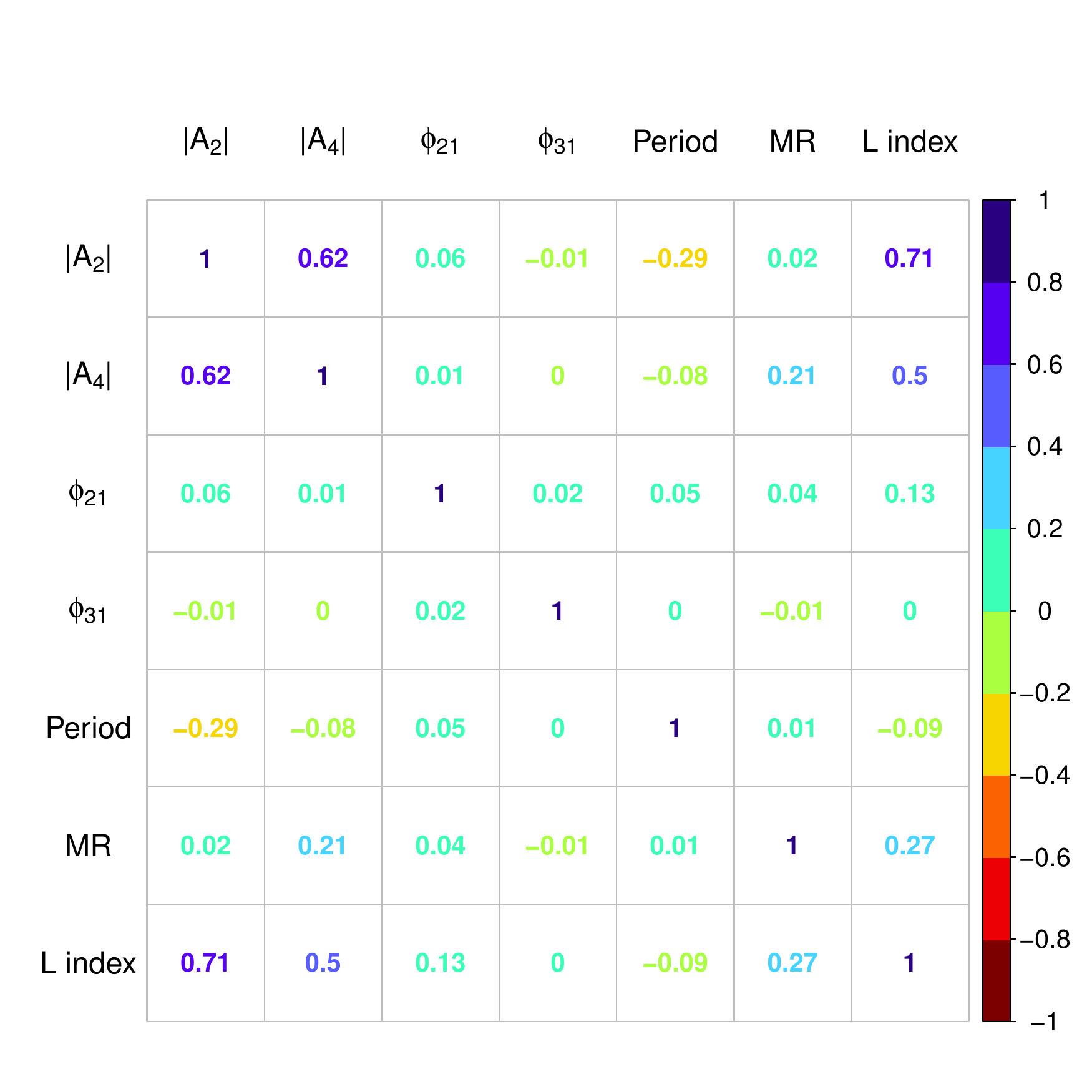}
\caption{Correlation matrix for our seven parameter feature vector.}
\label{fig:corrmat}
\end{center}
\end{figure}

Figure~\ref{fig:corrmat} shows the pair-wise correlation matrix where 
elements were computed using Pearson's linear correlation coefficient
$\rho_{ij}=\mbox{cov}(i,j)/(\sigma_i\sigma_j)$ for two features $i,j$
where cov is their sample covariance and $\sigma_i,\sigma_j$ are their sample
standard deviations. It's important to note that this only
quantifies the degree of {\it linear} dependency between any two
features. This is the type of dependency of interest since it
can immediately allow us to identify redundant features and hence reduce 
the dimensionality of our problem. The features that have the largest
correlation with any other feature are $|A_2|$, $|A_4|$, and $L$ index.
Although relatively high and significant (with a $<0.01\%$ chance of
being spurious), these correlations are still quite low 
compared to the typically recommended
value of $\rho\approx0.9$ at which to consider eliminating
a feature.

%\placefigure{fig:corrmat}

The pair-wise correlation matrix provided a first crude check, although
there could still be hidden collinearity in the data whereby one
or more features are captured by a linear combination of the
others. Our second test was therefore more general and
involved treating each feature in turn as the dependent variable
and testing if it could be predicted by any or all of the remaining
$M-1$ features (the independent variables in the regression fit).
We examined the $R^{2}$ values (coefficients of determination)
from each fit. These values
quantify the proportion of the variation in the dependent variable
that can be ``explained'' by some linear combination of all
the other variables (features). The highest $R^{2}$ value
was 0.68 and occured when $|A_2|$ was the dependent variable.
This was not high enough to warrant removing this feature.  
We also explored the fitted coefficients and their significance
from each linear fit. Most of them were {\it not} 
significant at the $<5\%$ level.
We conclude that none of the features exhibit sufficiently strong
correlations or linear dependencies to justify reducing
the dimensionality of our feature space.

\subsection{Training and Test Sample Preparation}\label{prep}

Before attempting to tune and train a RF classifier, we first partition
the input training sample described in section~\ref{tsamp} into two random
subsamples, containing 80\% and 20\% of the objects, or 6620 and 1653
objects respectively. These are respectively referred to as the
{\it true training sample} for use in tuning and training the RF
model using recursive cross-validation
(see below), and a {\it test sample} for performing a final validation check
and assessing classification performance by comparing known to predicted
outcomes. This {\it test sample} is sometimes referred to as the {\it hold-out}
sample. The reason for this random 80/20 split is to ensure that our
performance assessment (using the {\it test sample}) is independent of the
model development and tuning process (on the {\it true training sample}).
That is, we don't want to skew our performance metrics by a possibly
over-fitted model, however slight that may be.

\begin{figure}[hbtp]
\begin{center}
\includegraphics[scale=0.8]{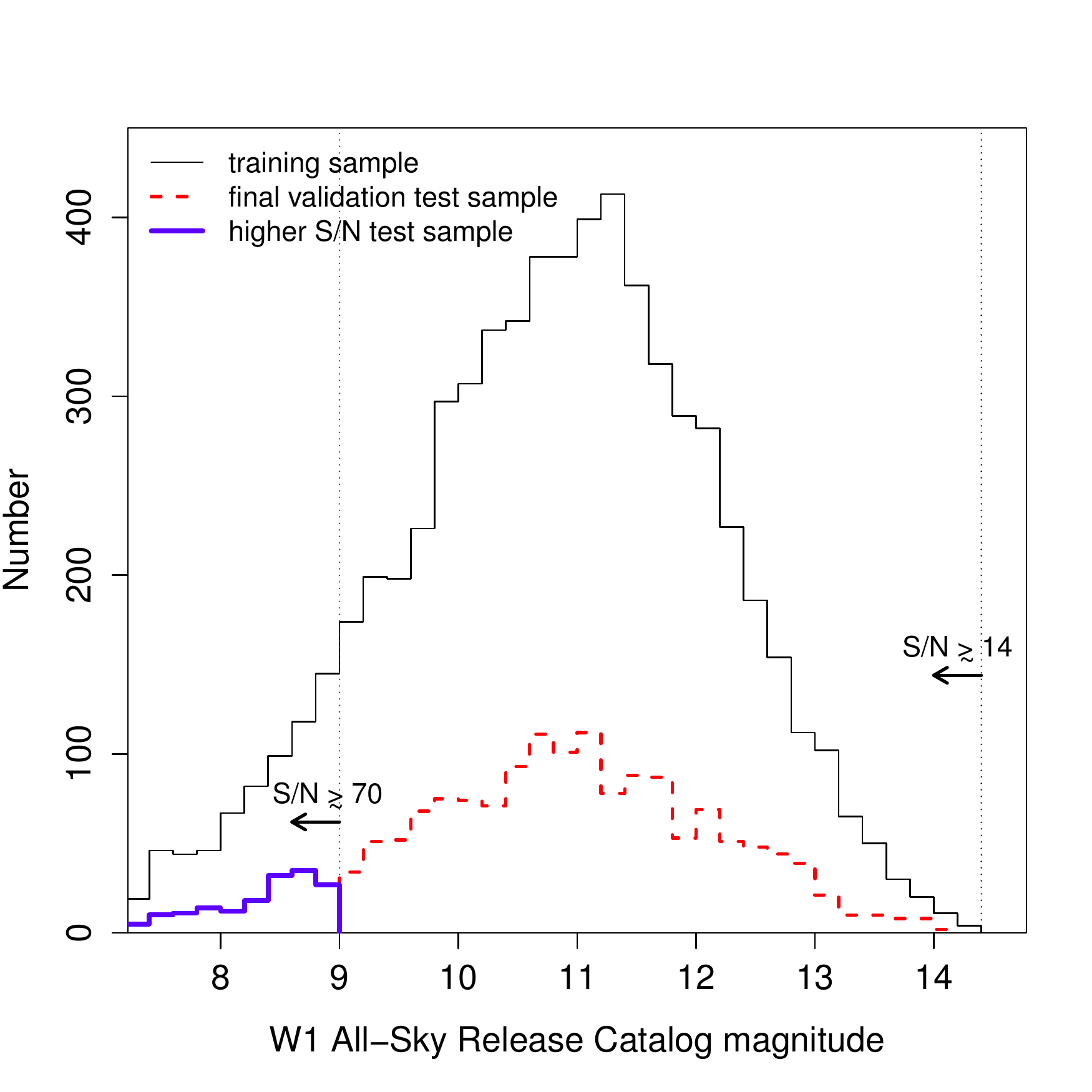}
\caption{W1 magnitude distributions from the WISE All-Sky Release Catalog
for all variables in our training sample, 
final test sample used for cross-validation, and a
brighter test subsample drawn from the final test sample using a magnitude
cut of 9. The Catalog magnitudes effectively represent the time-averaged
photometry from all single-exposure measurements.
The approximate signal-to-noise (S/N) ratios corresponding
to the limiting magnitudes of these samples are indicated.}
\label{fig:mags}
\end{center}
\end{figure}

In Figure~\ref{fig:mags}, we compare the W1 magnitude distributions for the
{\it true} training sample (referred to as simply ``training sample'' from
hereon), the test sample to support final validation, and from this,
an even smaller test subsample consisting of 194 objects with W1 magnitudes
$\leq 9$ mag. This bright subsample will be used to explore the classification
performance for objects with a relatively higher signal-to-noise (S/N) ratio.
The shape of the training and test-sample magnitude distributions
are essentially equivalent given both were randomly drawn from the same input
training sample as described above. That is, the ratio of the number of
objects per magnitude bin from each sample is approximately uniform within
Poisson errors. The magnitudes in Figure~\ref{fig:mags} are from the WISE
All-Sky Release Catalog \citep{cutri12} and derived from simultaneous 
Point Spread Function (PSF) fit photometry on the stacked single-exposures
covering each source from the 4-band cryogenic phase of the mission {\it only}.
These therefore effectively represent the time-averaged light curve photometry.
W1 saturation sets in at approximately 8 mag, although we included objects down
to 7 mag after examining the quality of their PSF-fit photometry and light
curves. The PSF-fit photometry was relatively immune to small amounts of
saturation in the cores of sources.

%\placefigure{fig:mags}

The S/N limits in Figure~\ref{fig:mags} are approximate and based on the
RMS-noise in repeated single-exposure photometry for the {\it non-variable}
source population \citep[section IV.3.b.ii. in][]{cutri12}.
These represent good overall proxies for the uncertainties in the light
curve measurements at a given magnitude. For example, the faintest sources
in our training sample have a time-averaged W1 magnitude of
$\approx 14.3$ mag where S/N $\approx 14$ and hence
$\sigma\approx1.086/14\approx0.08$ mag. This implies the fainter
variables need to have progressively larger variability amplitudes
in order to be reliably classified, e.g., with say
$\gtrsim 5\sigma$ or $\gtrsim0.4$ mag at W1 $\gtrsim14$ mag. This therefore
sets our effective sensitivity limit for detecting and characterizing
variability in the WISE single-exposure database. The paucity of
faint high-amplitude variables amongst the {\it known} variable-source
population in general explains the gradual
drop-off in numbers beyond W1 $\approx11.3$ mag.

\subsection{Training and Tuning the Random Forest Classifier}\label{train}

An overview of the RF machine learning algorithm was given in 
section~\ref{rf}. Even though the RF method is referred to as
a {\it non-parameteric} classification method, it still has
a number of tuning parameters to control its flexibility.  
In this paper, these are (1) the number of decision trees $N_{\rm tree}$
to build from each boostrapped sample of the
training set; (2) the number of features $m_{\rm try}$ to randomly
select from the full set of $M$ features to use as  
candidates for splitting at each tree node; and (3) the  
size of a terminal node in the tree, {\it nodesize},
represented as the minimum number of objects allowed in
the final subregion where no more splitting occurs.
For classification problems (as opposed to regression where the
response is a multi-valued step function), \citet{breiman01} recommends
building each individual tree right down to its leaves where
{\it nodesize} $= 1$, i.e., leaving the trees ``unpruned''. 
This leaves us with $N_{\rm tree}$ and $m_{\rm try}$. The optimal
choice of these parameters depends on the complexity of the
classification boundaries in the high-dimensional feature space. 

We formed a grid of $N_{\rm tree}$ and $m_{\rm try}$ test values and our 
criterion for optimality (or figure-of-merit) was chosen to be the
{\it average} classification accuracy. This is defined as the ratio of
the number of correctly predicted classifications from the specific
RF model to the total number of objects in all classes. 
This metric is also referred to as the average classification efficiency
and ``1 - accuracy'' is the error rate. This is a suitable metric to use
here since our classes contain similar numbers of objects and the overall
average accuracy will not be skewed towards the outcome for any
particular class. This metric would be biased for instance if one class was
substantially larger than the others since it would dictate the average
classification accuracy.

Fortunately, the classification accuracy is relatively insensitive
to $N_{\rm tree}$ when $N_{\rm tree}$ is large ($>$ a few hundred) and
$m_{\rm try}$ is close to its optimum value.
The only requirement is that $N_{\rm tree}$ be large
enough to provide good averaging (``bagging'') to minimize
the tree-to-tree variance and bring
the prediction accuracy to a stable level, but not too
large as to consume unnecessary compute runtime.
Therefore, we first fixed $N_{\rm tree}$ at a 
relatively large value of 1000, then
tuned $m_{\rm try}$ using a 10-fold cross-validation on the 
{\it true} training sample defined in section~\ref{prep} and selecting the 
$m_{\rm try}$ value that maximized the classification accuracy (see below).
Once an optimal value of $m_{\rm try}$ was found, we then explored
the average classification accuracy as a function of $N_{\rm tree}$
to select an acceptable value of $N_{\rm tree}$.
10-fold cross-validation (or $K$-fold in general) entails  
partitioning the training sample into ten subsamples where 
each subsample is labeled $k = 1,2,3...10$, then training the RF model
on nine combined subsamples and predicting classifications for the
remaining one. These predictions are compared to the known (true)
classifications to assess classification performance. This prediction
subsample is sometimes referred to as the ``hold-out'' or 
``out-of-bag'' sample.
Given $N$ objects in the training sample, we iterate until every subsample $k$ 
containing $N/10$ objects has served as the prediction 
dataset using the model trained on the remaining $T=9N/10$ objects. 
The final classification (or prediction) performance is then 
the average of all
classification accuracies from all 10 iterations.

The \texttt{caret} package in \texttt{R} provides a
convenient interface to train and fit a RF model using $K$-fold
cross-validation. This calls the higher level {\it randomForest}() function,
an implementation of the original \citet{breiman04} algorithm written
in Fortran. We were unable to find a precise description of
the algorithm implemented in tandem with cross-validation
by the \texttt{R} \texttt{caret} package.
Given that there is a lot happening in the training and tuning
phase, we lay out the steps in Appendix~\ref{appA}.

\begin{figure}[hbtp]
\begin{center}
\includegraphics[scale=0.8]{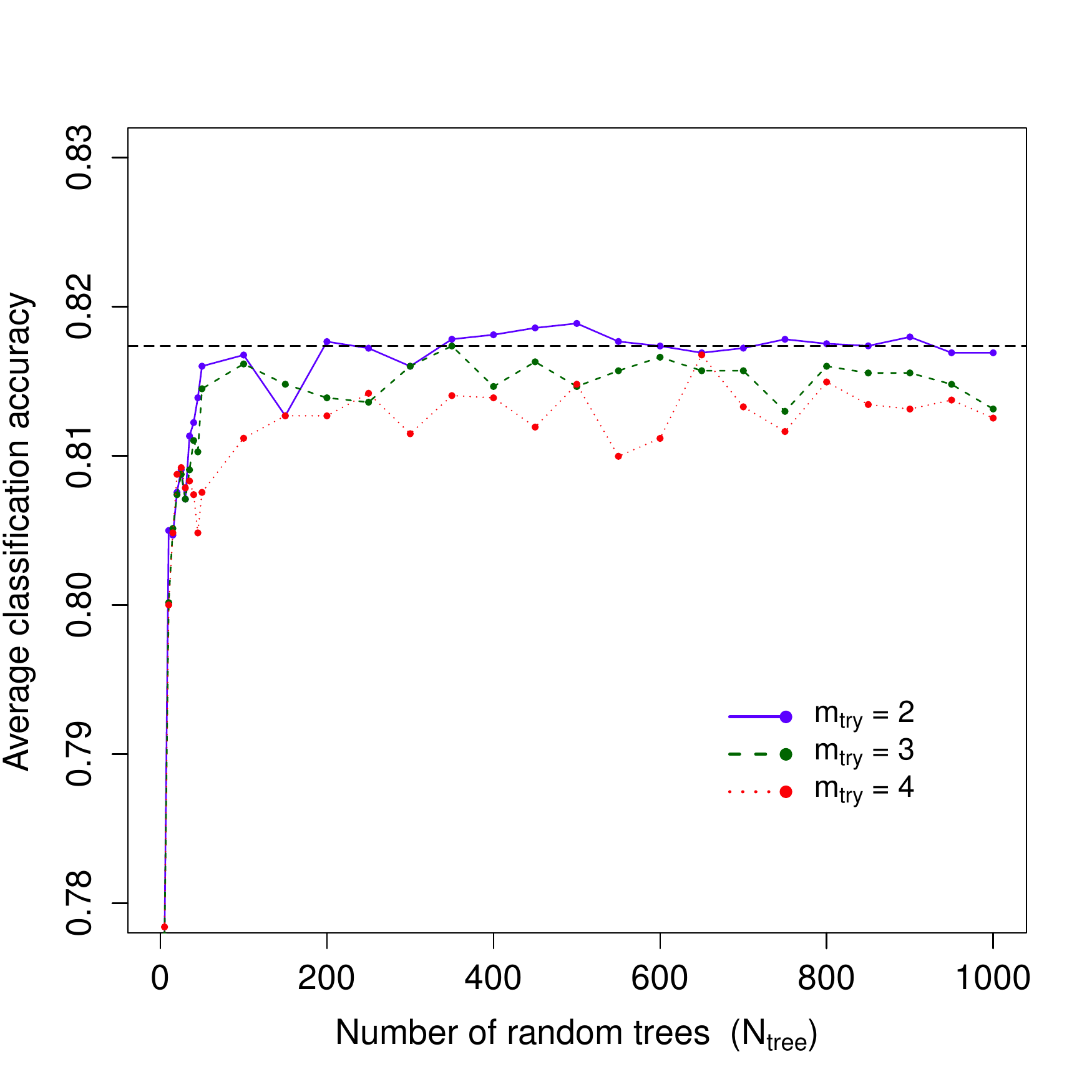}
\caption{Average classification accuracy from cross-validation as a
function of the number of randomly generated trees in the random forest
(i.e., the $N_{\rm tree}$ parameter) for three values of $m_{\rm try}$.
The horizontal dashed line denotes the asymptotic value of the
classification accuracy ($\approx 0.817$) for the optimal
value $m_{\rm try} = 2$.}
\label{fig:ntree}
\end{center}
\end{figure}

Figure~\ref{fig:mtry} shows the average classification accuracy as a function
of the trial values of $m_{\rm try}$. The optimal value is
$m_{\rm try} = 2$ and close to the rule-of-thumb suggested by
\citet{breiman01}: $m_{\rm try}\approx\sqrt{M}$, where $M$ is the
total number of features (7 here). As mentioned earlier,
the classification accuracy is relatively insensitive
to $N_{\rm tree}$ when $N_{\rm tree}$ is large and
$m_{\rm try}$ is close to its optimum value.
The results in Figure~\ref{fig:mtry} assume a fixed value $N_{\rm tree}=1000$.
Figure~\ref{fig:ntree} shows the average classification accuracy as a function 
of $N_{\rm tree}$ for $m_{\rm try} =$ 2, 3, and 4. The achieved accuracies are
indeed independent of $N_{\rm tree}$ for $N_{\rm tree}\gtrsim 400$. 
However, to provide good tree-averaging (``bagging'') and hence keep
the variance in the final RF model fit as small as possible,
we decided to fix $N_{\rm tree}$ at 700. This also kept the compute runtime
at a manageable level.

%\placefigure{fig:mtry}

%\placefigure{fig:ntree}

When predicting the classification for
a new object with feature vector $\mathbf{X}$,
it is pushed down the tree. That is, it is assigned the
label of the training sample in the terminal node it ends up in. This 
procedure is iterated over all $N_{\rm tree}$ trees in the 
ensemble, and the mode (or majority) vote of all trees is
reported as the predicted class. However, instead of the winning class,
one may want to quote the probabilities that an object belongs
to each respective class. This allows one to make a more informed decision.
The probability that a new object with feature vector
$\mathbf{X}$ belongs to some class $C_j$ where $j = 1,2,3,...$
is given by
\begin{equation}\label{eq:prob}
P\left(\mathbf{X}|C_j\right) = \frac{1}{N_{\rm tree}}
      \sum_{i=1}^{N_{\rm tree}}I\left(p_i=C_j|\mathbf{X}\right),
\end{equation}
where $I(p_i=C_j|\mathbf{X})$ is an indicator function defined to
be 1 if tree $i$ predicts class $C_j$ and 0 otherwise.
In other words, for the RF classifier, the class probability is simply
the fraction of $N_{\rm tree}$ trees that predicted that class. 
It's important to note that the probability computed via equation~\ref{eq:prob} 
is a {\it conditional} class probability and only has meaning {\it relative}
to the probabilities of obtaining the same
features $\mathbf{X}$ conditioned
on the other contending classes. That is, we say an object with features
$\mathbf{X}$ is relatively more likely to have been generated by the
population of objects defined by class $C_1$ than classes $C_2$ or $C_3$, etc.  

The definition in equation~\ref{eq:prob} should not be confused 
with the {\it posterior} probability
that the object belongs to class $C_j$ in an ``absolute'' sense, i.e., as
may be inferred using a Bayesian approach. This can be done by assuming some 
{\it prior} probability $P(C_j)$ 
derived for example from the proportion that each class contributes
to the {\it observable} population of variable sources. The probability in this
case would be:
\begin{equation}\label{eq:bprob}
P\left(C_j|\mathbf{X}\right) = \frac{P\left(\mathbf{X}|C_j\right)
      P\left(C_j\right)}{P\left(\mathbf{X}\right)},
\end{equation}
where $P(\mathbf{X})$ is the normalization factor that ensures the integral of
$P(C_j|\mathbf{X})$ over all $C_j$ is 1
and $P(\mathbf{X}|C_j)$, the ``likelihood'', is
given by equation~\ref{eq:prob}.
Unfortunately, plausible values for the priors $P(C_j)$ are difficult 
to derive at this time since
the relative number of objects across classes in our training sample are 
likely to be subject to heavy selection biases, e.g., brought about by both
the WISE observational constraints and heterogeneity of the input
optical variability catalogs used for the cross-matching. The current
relative proportions of objects will not represent the true mix of variable
sources one would observe in a controlled flux-limited sample according to 
the WISE selection criteria and sensitivity to each class.
This Bayesian approach will be considered
in future as classification statistics and selection effects are
better understood. For our initial classifications, we will quote the  
relative (conditional) class probabilites defined by equation~\ref{eq:prob}
(see section~\ref{cpred} for details).

\subsection{Quantifying Feature Importance}\label{import}

An initial qualitative assessment of the separability of our three broad
variability classes according to the seven mid-IR light curve features
was explored in Section~\ref{fan}. Given the relatively high
dimensionality of our feature space, this separability is difficult
to ascertain by looking at pairwise relationships alone. Our goal is
to explore class separability using the full feature space in more detail.
This also allows us to identify those features that best discriminate each
class as well as those that carry no significant information, both
overall and on a per-class basis.

{\it Random forests} provide a powerful mechanism to measure the predictive
strength of each feature for each class, referred generically to as
{\it feature importance}. This quantifies, in a relative sense, the impact
on the classification accuracy from
randomly permuting a feature's values, or equivalently, forcing a feature
to provide no information, rather than what it may provide on input.
This metric allows one to determine which features work best at
distinguishing between classes and those that physically define each class.
Feature importance metrics can be additionally generated in the
RF training/tuning phase using the ``hold-out'' (or ``out-of-bag'') subsamples
during the cross-validation process (Section~\ref{train}),
i.e., excluded from the bootstrapped training samples.
The prediction accuracies from the {\it non-permuted}-feature and
{\it permuted}-feature runs are differenced then averaged over all the
$N_{\rm tree}$ trees and normalized by their standard-error
($\hat{\sigma}/\sqrt{N_{\rm tree}}$). The importance metrics
are then placed on a scale of 0 - 100 where the
``most important'' metric is assigned a value of 100 for the class where
it is a maximum. The intent here is that features that lead to
large differences in the classification accuracy for a specific class
when their values are randomly permuted are also likely to be more
important for that class. It's worth noting that even though this
metric is very good at finding the {\it most important} features, it can give
misleading results for highly-correlated features which one 
might think are important.
A feature could be assigned a low RF importance score (i.e., with little
change in the cross-validation accuracy after
its values are permuted) simply because other features that strongly correlate
with it will stand in as ``surrogates'' and carry its predictive power. 

\begin{figure}[hbtp]
\begin{center}
\includegraphics[scale=1.0]{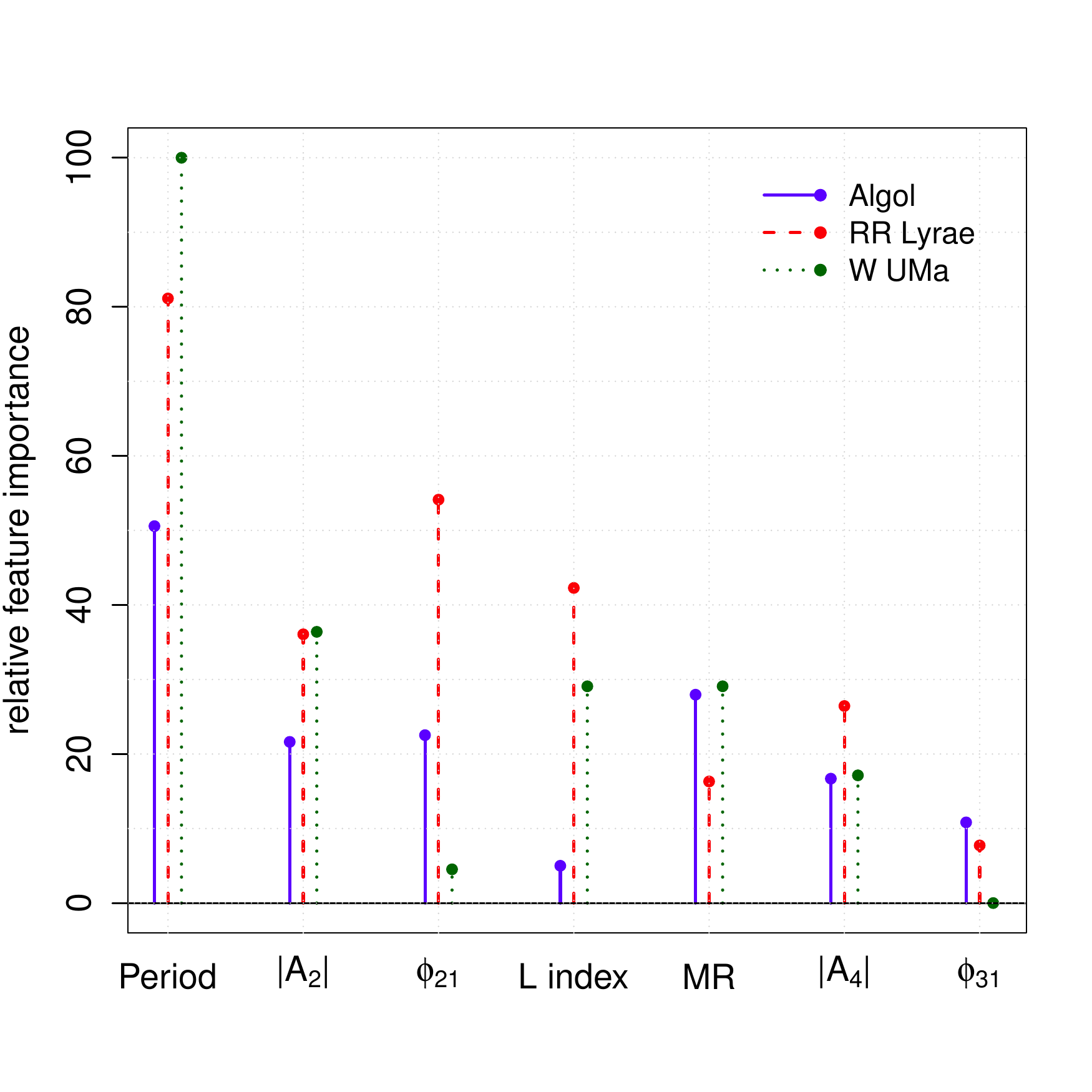}
\caption{Relative importance of each light curve feature for each
of the three classes, where a higher ``importance value'' implies
the feature is better at discriminating and predicting
a particular class using the RF classifier. The features are shown in 
order of decreasing {\it average} relative importance across classes 
(left to right).}
\label{fig:import}
\end{center}
\end{figure}

Figure~\ref{fig:import} illustrates the relative importance of each
feature for each class using the predictions from our initially tuned RF
classifier. Period is the most important feature for all classes. This
is no surprise since the three classes are observed to occupy almost distinct
ranges in their {\it recovered} periods (Figure~\ref{fig:period}),
therefore providing good discriminative and predictive power.
In general, not all features are equally as important across classes.
For example, the relative phase $\phi_{21}$ and $L$-index are relatively
weak predictors on their own for the W Uma and Algol eclipsing binaries
respectively while they are relatively strong predictors for RR Lyr pulsating
variables. In practice, one would eliminate the useless features (that carry
no information) across all classes and retrain the classifier. However, since
all features have significant predictive power for at least one class
in this study, we decided to retain all features.

%\placefigure{fig:import}

We explored the impact on the {\it per-class} prediction accuracy of randomly
permuting the $\phi_{21}$ and $L$-index feature values, i.e., so they provide
no useful information. The RF classifier was retrained and the final validation
{\it test sample} defined in Section~\ref{prep} was used to assess the
classification accuracy. This provided more of an ``absolute''
measure of the importance of these features than in Figure~\ref{fig:import}.
We found that if either
$\phi_{21}$ or $L$-index were forced to be useless, the classification
accuracies for the Algol and W Uma classes were not significantly altered.
However, the accuracy dropped by $\approx 4.1$\% and $\approx 3.1$\%
for the RR Lyr class by forcing these features to be useless respectively.
If both $\phi_{21}$ and $L$-index were forced to be useless, the change in
classification accuracies for the Algol and W Uma classes were still
insignificant (dropping by $\lesssim 1.5$\%), but the drop for the
RR Lyr class was $\approx 7.7$\%. This not only confirms the results in
Figure~\ref{fig:import}, but also the fact that RF classifiers are
relatively immune to features that carry little or no class information.

\subsection{Classifier Performance using Cross-Validation}\label{cv}

\begin{figure}[hbtp]
\begin{center}
\includegraphics[scale=0.8]{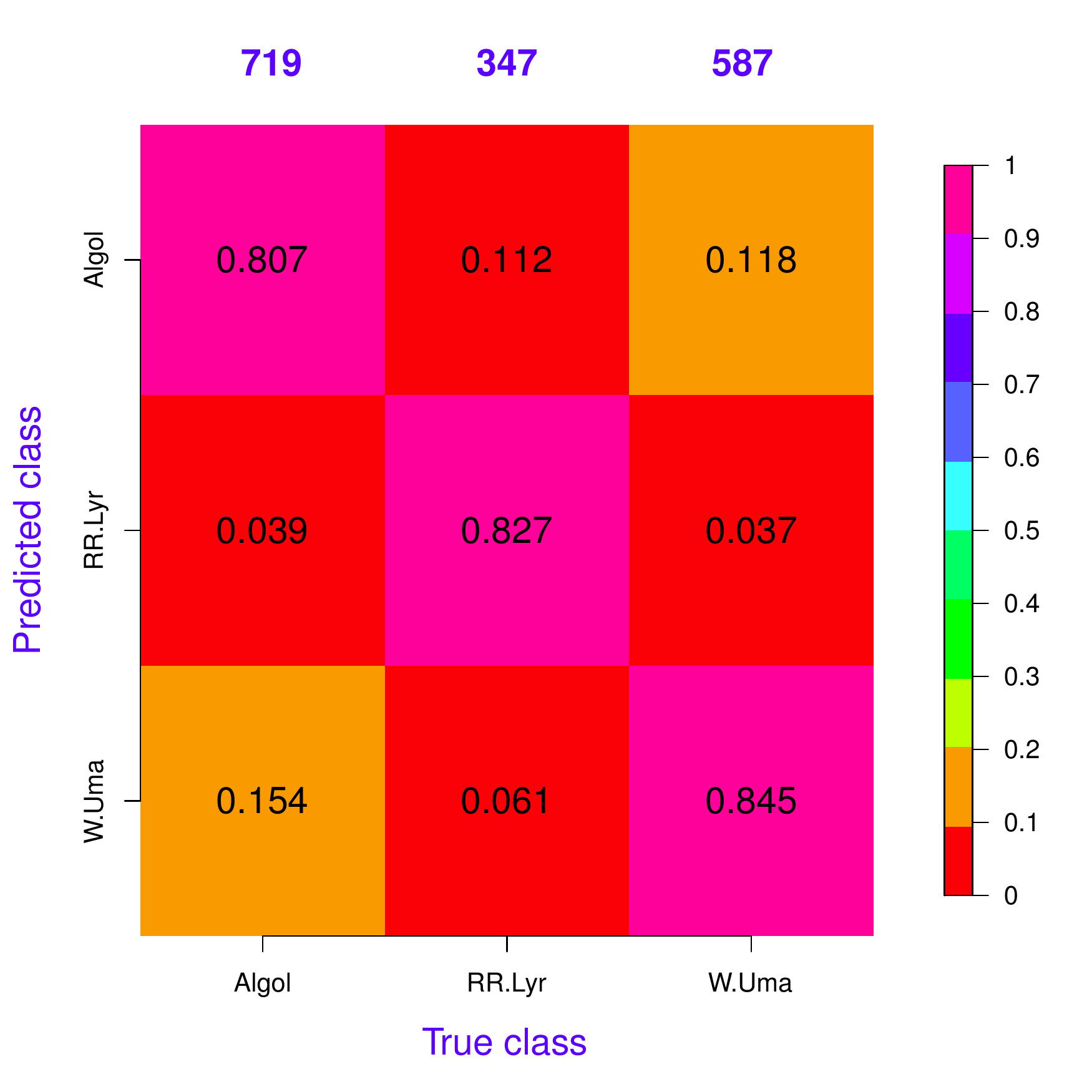}
\caption{Confusion matrix for our three classes of interest using our
largest {\it test sample}. Classification accuracies (or efficiences)
are along the diagonal. A perfect classifier would place all mass
on the diagonal. The numbers above the matrix are the {\it true} number
of objects in each class. See Section~\ref{cv} for details.}
\label{fig:cv}
\end{center}
\end{figure}

We validate the overall accuracy of the RF classifier that was fit to the
training sample by predicting classifications for the two {\it test samples}
defined in Section~\ref{prep} (Figure~\ref{fig:mags}) and comparing
these to their ``true'' (known) classifications. These test samples
are independent of the training sample and hence allow an unbiased
assessment of classifier performance. This was explored by computing
the {\it confusion matrix} across all classes. The confusion matrix
for our largest test sample (consisting of 1653 objects to W1$\sim14$ mag)
is shown in Figure~\ref{fig:cv}. The quantities therein represent the
proportion of objects in each true (known) class that were predicted
to belong to each respective class, including itself. The columns are
normalized to add to unity. When compared
to itself (i.e., a quantity along the diagonal going from
top-left to bottom-right in Figure~\ref{fig:cv}), it is referred to as
the {\it sensitivity} in machine learning parlance. It is also loosely
referred to as the per-class {\it classification accuracy}, {\it efficiency}
(e.g., as in Section~\ref{import}), or {\it true positive rate}. We obtain
classification efficiencies of 80.7\%, 82.7\%, and 84.5\% for
Algols, RR Lyrae, and W Uma type variables respectively. The overall
classification efficiency, defined as the proportion of all objects
that were correctly predicted (irrespective of class) is
$\approx82.5$\%. The corresponding 95\% confidence interval 
(from bootstrapping) is 80.5\% to 84.3\%, or approximately
$\pm$2\% across all three classes.

%\placefigure{fig:cv}

For comparison, \citet{richards11} obtained
an overall classification efficiency of $\approx77.2$\% on a
25-class dataset of 1542 variable stars from the OGLE and
Hipparcos surveys. However, if we isolate their Algol
(predominately $\beta$ Lyrae), RR Lyrae
(all types) and W Uma statistics, we infer an overall classification
efficiency of $\approx88.4$\%, implying an improvement
of $\approx6\%$ over our estimate for WISE variables.
This difference is likely due to their higher quality, longer
timespan optical light-curves -- specially selected to have
been well studied in the first place. Nonetheless, our classification
performance is still good given the WISE cadence,
sparsity and timespan of observations, and possible uncertainties in
classifications from the literature used to validate the
predictions.

The off-diagonal quantities of the confusion matrix in Figure~\ref{fig:cv}
can be used to compute the reliability (or equivalently the
{\it purity}, {\it specificity}, or ``$1-$ {\it false positive rate}'')
for a specific class. That is, the proportion of objects in all
{\it other} classes that are {\it correctly} predicted to {\it not} contaminate
the class of interest. This can be understood by noting that the
only source of unreliability (or contamination) to each class
are objects from other classes. For example, the false positive rate
($FPR$) for the Algol class is
\begin{eqnarray*}\label{eq:fpr}
FPR & = & \frac{0.112\times347 + 0.118\times587}
               {347\times\left(0.112+0.827+0.061\right)+
                587\times\left(0.118+0.037+0.845\right)} \\
    & \approx & 0.116
\end{eqnarray*}
and hence its purity is $1 - FPR\approx88.4$\%. Similarly, the purity
levels for the RR Lyrae and W Uma classes from Figure~\ref{fig:cv} are
96.2\% and 87.6\% respectively. For comparison, \citet{richards11} obtain
purity levels of up to 95\% for these and most other classes in their study.

For the smaller, higher S/N test sample of 194 objects with W1 magnitudes
$\leq 9$ (Figure~\ref{fig:mags}), the classification accuracy for the
Algol class improves to $\approx89.7$\%, compared to
80.7\% for the large test sample. However for the RR Lyr class, the
classification accuracy drops to 55.5\% (from 82.7\%) and for 
the W Uma class, it drops to 79.3\% (from 84.5\%). In general, we would
have expected an increase in classification accuracy across all classes
when only higher S/N measurements, and hence objects with
more accurately determined features are used.
This indeed is true for the Algol class which appears to be the most
populous in this subsample with 127 objects. The drop in classification
performance for the other two classes can be understood by low number
statistics with only 9 RR Lyr and 58 W Uma objects contributing.
Their sampling of the feature space density distribution in the training set
for their class is simply too sparse to enable reliable classification metrics
to be computed using ensemble statistics on the predicted outcomes.
In other words, there is no guarantee that most of the 9 RR Lyr in this high
S/N test sample would fall in the densest regions of the RR Lyr training
model feature space so that they can be assigned high enough
RF probabilities to be classified as RR Lyr.

\begin{figure}
\begin{minipage}[t]{0.45\textwidth}
\includegraphics[scale=0.35]{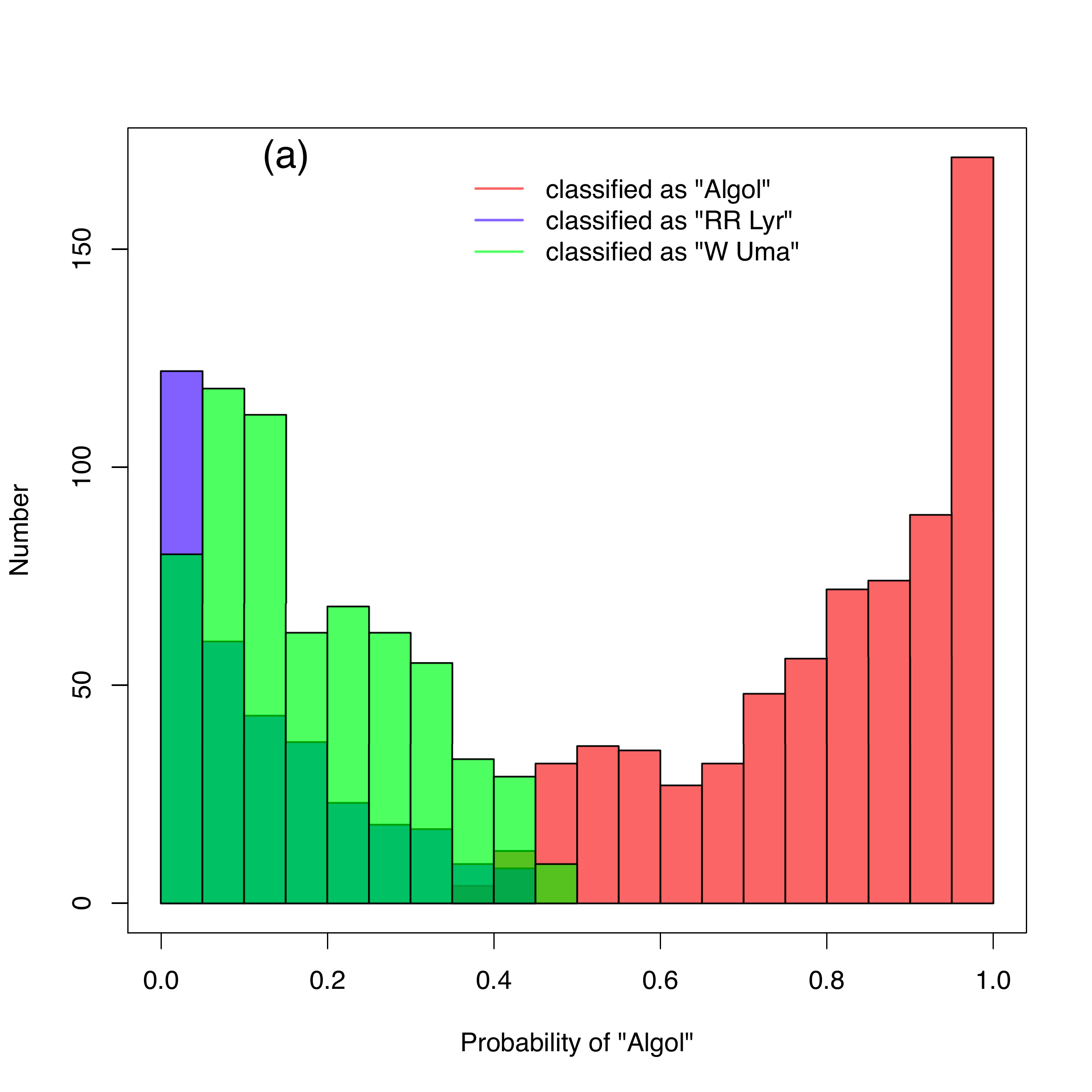}
\end{minipage}
\hspace{\fill}
\begin{minipage}[t]{0.45\textwidth}
\includegraphics[scale=0.35]{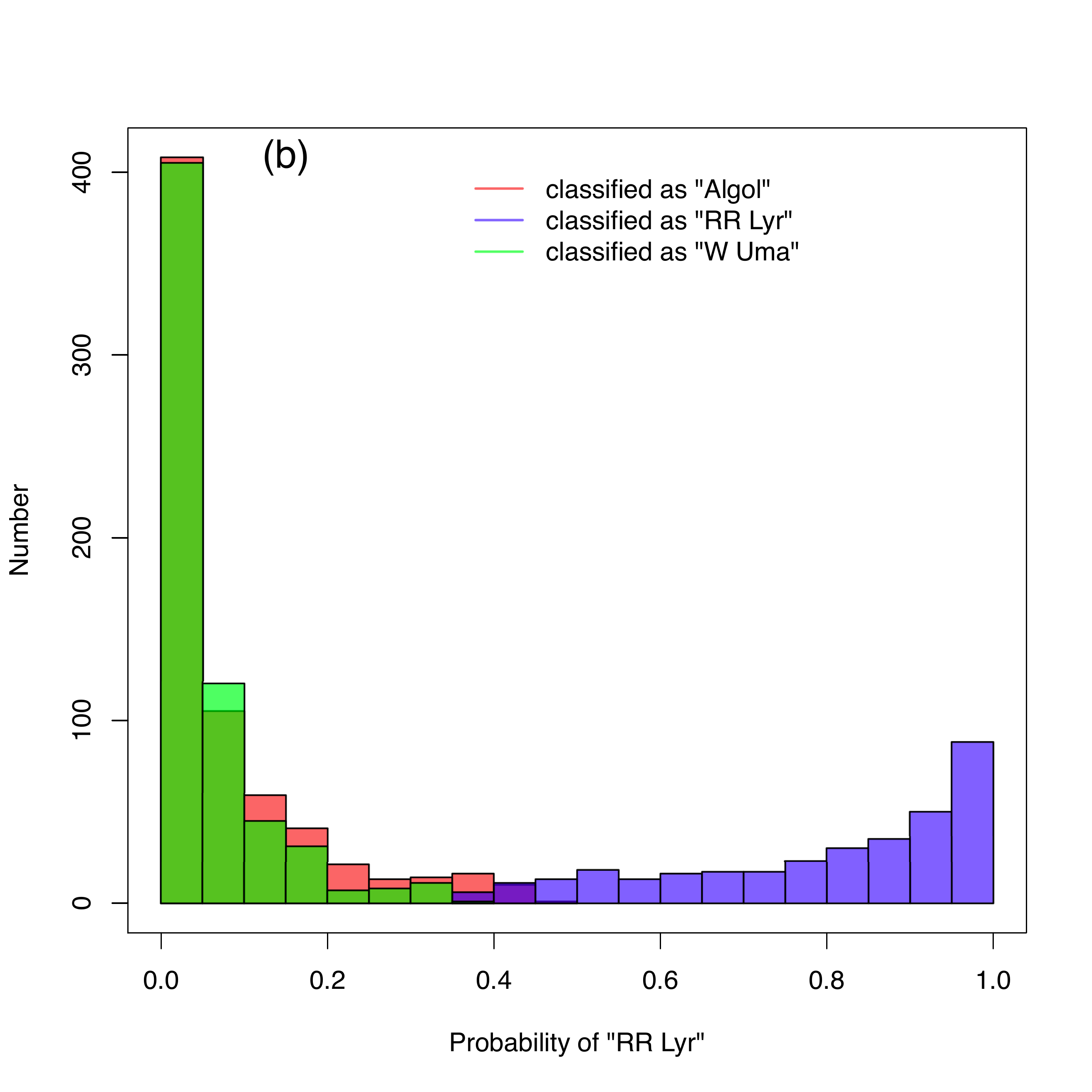}
\end{minipage}
\vspace*{0.5cm}
\begin{minipage}[t]{0.45\textwidth}
\includegraphics[scale=0.35]{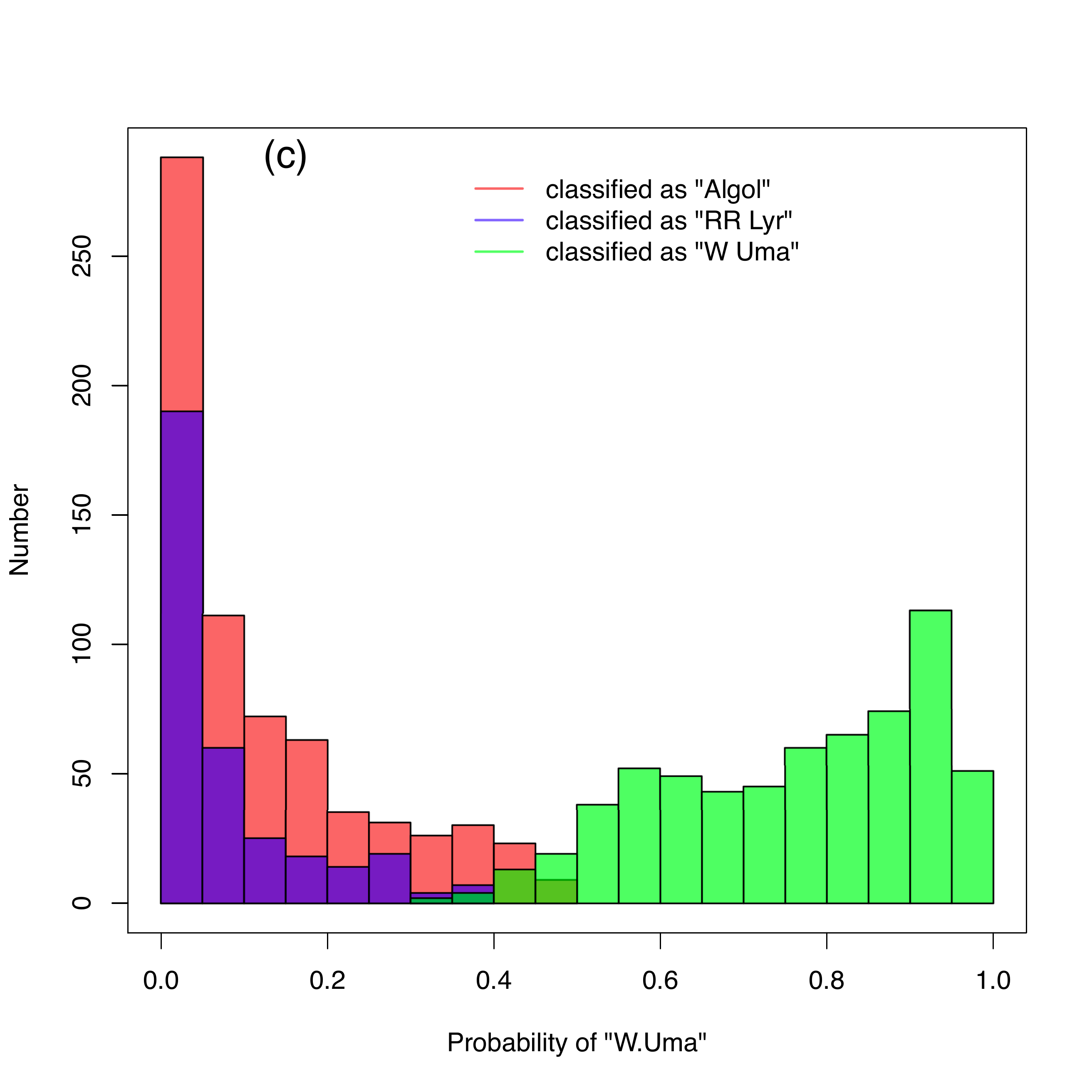}
\end{minipage}
\caption{Histograms of classification probabilities from the RF method for
the three classes: (a) Algol; (b) RR Lyrae; and (c) W Uma
variables conditioned on the ``winning'' class assigned by the RF method
(color coded). The various color shades are where the histograms from
different classes overlap.}
\label{fig:rfprob}
\end{figure}

The primary output from the RF classifier when predicting the outcome
for an object with given features is a vector of {\it conditional class} 
likelihoods as defined by equation~\ref{eq:prob}. By default, the 
``winning'' class is that with the highest likelihood.
A more reliable and secure scheme to assign the winning class will be 
described in Section~\ref{cpred}. 
Distributions of all the classification probabilities for our
largest test sample are shown in Figure~\ref{fig:rfprob}.
These probabilities are conditioned on the winning class that was
assigned by the RF classifier so that histograms at the high
end of the probability range in each class-specific panel 
correspond to objects in that winning class.
The spread in winning class probabilities is similar across the
three classes, although the Algol class has slightly more mass
at $P\gtrsim 0.7$ (Figure~\ref{fig:rfprob}a).
This indicates that the 7-D feature space 
sample density is more concentrated (or localized) for this class than 
for the other classes.

%\placefigure{fig:rfprob}

\begin{figure}[hbtp]
\begin{center}
\includegraphics[scale=0.8]{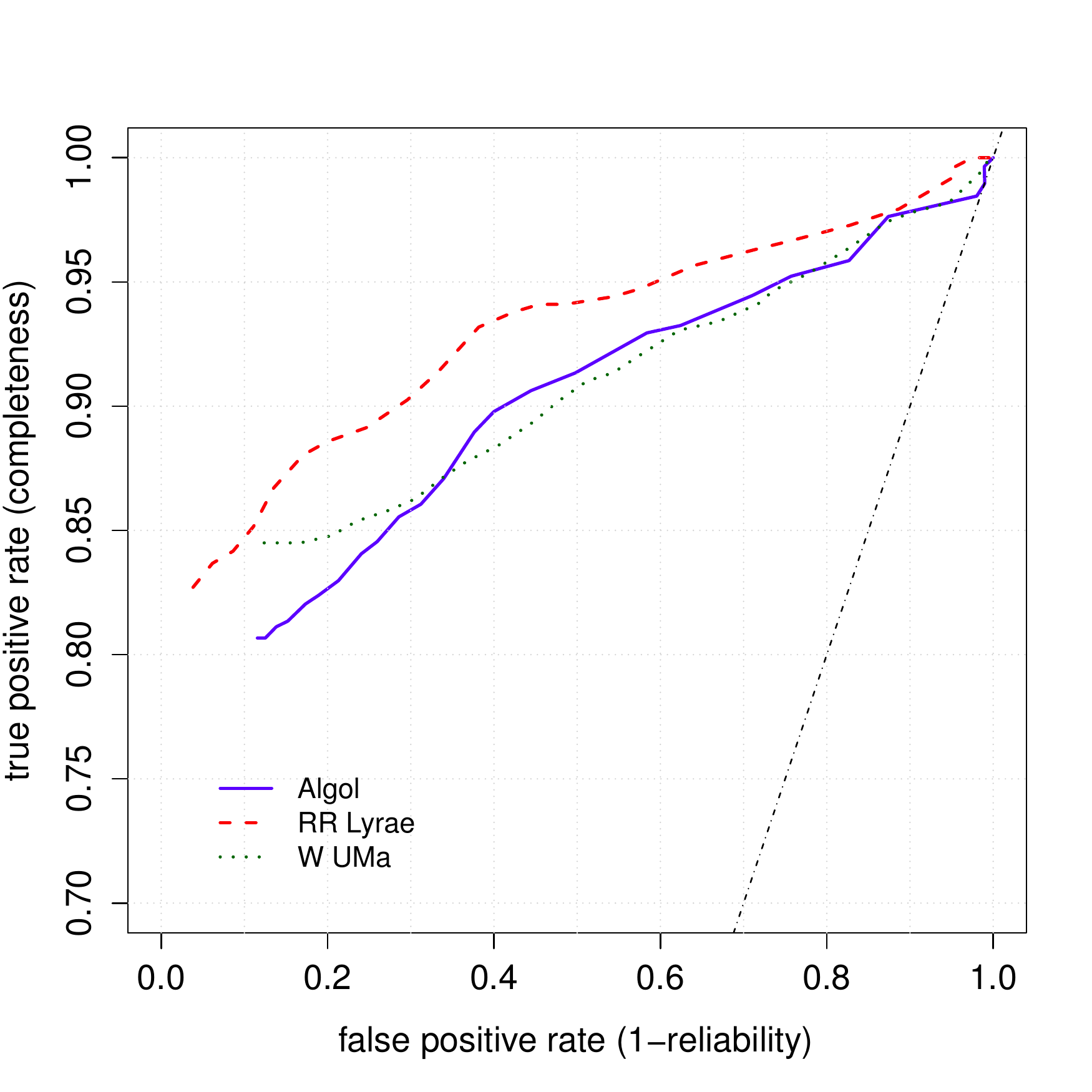}
\caption{Receiver operating characteristic (or ROC) curves for each target
class (color coded) by thresholding their classification probabilities.
The lowest thresholds are at the far left and the highest ($prob > 0.98$)
are at the far right. The black dot-dashed line is the line of 
equality ($TPR = FPR$) and represents the result from randomly assigned 
classifications with points above it being better than random.}
\label{fig:roc}
\end{center}
\end{figure}

Figure~\ref{fig:roc} shows the receiver operating characteristic (ROC) 
curves for each class in our largest test sample. These are generated by
thresholding on the classification probabilities of objects in each
class (i.e., with $P > 0, P > 0.02, P > 0.04, ..., P > 0.98$ 
from {\it left to right} in Figure~\ref{fig:roc}), 
then computing the confusion matrix
for each thresholded subclass. The {\it true positive rate}
($TPR$ or classification accuracy)
and {\it false positive rate} ($FPR$ or impurity) were then extracted to create
the ROC curve. The trends for these curves are as expected.
Given the class probability quantifies the degree of
confidence that an object belongs to that class,
the larger number of objects sampled to a lower {\it cumulative}
probability level will reduce both the overall $TPR$ and $FPR$.
That is, a smaller fraction of the truth is recovered, but
the number of contaminating objects (false positives) from other 
classes does not increase much and the larger number of objects
in general keeps the $FPR$ relatively low.
The situation reverses when only objects with a higher classification
probability are considered.
In this case there are fewer objects in total 
and most of them agree with the true class (higher $TPR$).
However, the number of contaminants is not significantly lower (or does
not decrease in proportion to the reduced number of objects) and hence
the $FPR$ is slightly higher overall.

%\placefigure{fig:roc}

It is also interesting to note that even though the {\it full} test sample
confusion matrix in Figure~\ref{fig:cv} indicates that W Uma objects have the
highest classification accuracy (at 84.5\% -- corresponding to far left
on the ROC curve in Figure~\ref{fig:roc}), this is overtaken by
RR Lyrae at internal probability thresholds of $P\gtrsim0.1$ where the
classification accuracy ($TPR$) becomes $>86\%$. This however is at the 
expense of an increase in the $FPR$ to $>12\%$.
Therefore, the ROC curves contain useful information for selecting
(class-dependent) classification probability thresholds such that
specific requirements on the $TPR$ and $FPR$ can be met. 

\subsection{Comparison to other Classifiers}

We compare the performance of the RF classifier trained above to other
popular machine-learned classifiers. The motive is to provide a cross-check
using the same training data and validation test samples. We explored
artificial neural networks (NNET), $k$-Nearest Neighbors ($k$NN),
and support vector machines (SVM). A description of these methods can
be found in \citet{hastie09}. The \texttt{R} \texttt{caret} package
contains convenient interfaces and functions to train, tune, test, and compare
these methods \citep{kuhn08}. More ML methods are available, but
we found these four to be the simplest to set-up and tune for our problem
at hand. Furthermore, these methods use very dissimilar algorithms and
thus provide a good comparison set. Parameter tuning was
first performed automatically
using large grids of test parameters in a 10-fold cross-validation
(defined in Section~\ref{train}); then the parameter
ranges were narrowed down
to their optimal ranges for each method
using grid sizes that made the effective number
of computations in training approximately equal across methods.
This enabled a fair comparison in training runtimes.

%\placetable{comp}

\begin{figure}[hbtp]
\begin{center}
\includegraphics[scale=0.8]{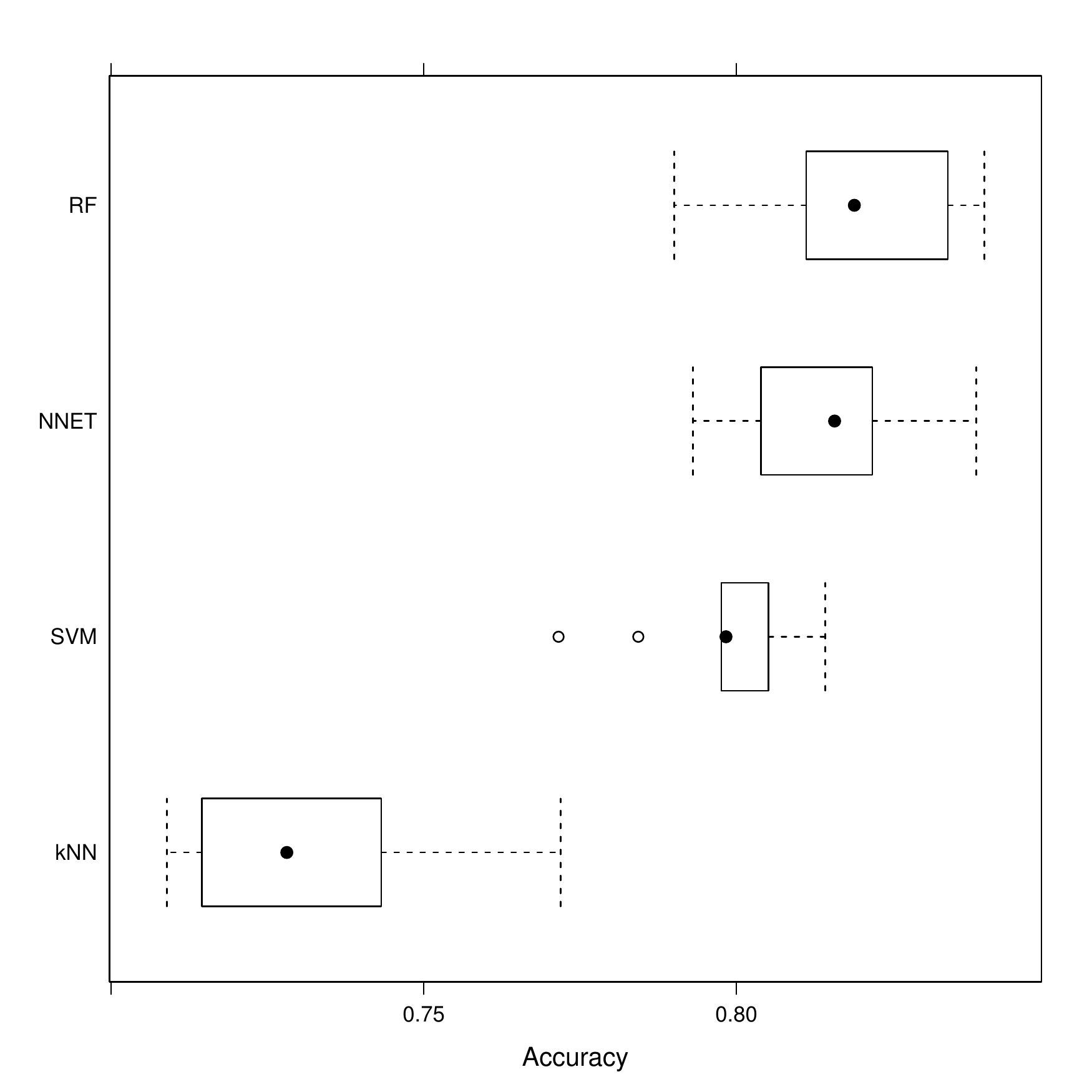}
\caption{Distribution of average classification accuracies (or efficiencies)
from cross-validation for four
machine-learning methods represented as box and
whisker diagrams. The filled circles are medians; the boundaries
of the central boxes represent interquartile ranges (25 to 75\%); the
outer whiskers define the boundaries for outliers (1.5 times the
interquartile range below the first quartile or above the
third quartile), and open circles are outliers.}
\label{fig:cfbox}
\end{center}
\end{figure}

The classification
accuracies (or efficiencies), runtimes, and the significance of the
difference in mean accuracy relative to the RF method are
compared in Table~\ref{comp}. The latter is in terms of the $p$-value
of obtaining an observed difference of zero (the null hypothesis)
by chance according to a paired $t$-test.
It appears that the NNET method performs just as well as the RF method
in terms of classification accuracy, although the RF method has
a slight edge above the others. This can be seen in
Figure~\ref{fig:cfbox} where the overall distributions in accuracy
are compared. Aside from the similarity in classification performance between
NNET and RF, the added benefits of the RF method, e.g.,
robustness to outliers, flexibility and ability to capture
complex structure, interpretability, relative immunity to irrelevent and
redundant information, and simple algorithms to measure feature importance and
proximity for supporting active learning frameworks (Section~\ref{al}),
makes RF our method of choice.

%\placefigure{fig:cfbox}

\section{Constructing the WVSDB and Assigning Classifications}\label{cpred}

Our goal for the WVSDB is to report all {\it periodic} variable star
types as allowed by the WISE observing constraints using the best quality
photometric time-series data from the primary-mission (cryogenic and
post-cryogenic) single-exposure Source Databases.
Candidate variables will be selected using a relatively high value of the
WISE Source Catalog variability flag ($var\_flg\geq 6$). Recently,
$var\_flag$ was made more reliable compared to the version initially
used to construct our training sample (Section~\ref{tsamp}).
The new $var\_flag$ is included in the recently released AllWISE
Source Catalog \citep[][section V.3.b.vi]{cutri13} and is based on a
combination of metrics derived directly from the single-exposure flux
time-series. This includes the significance of correlated variability in
the W1 and W2 bands. In addition, candidates will be selected using other
quality and reliability metrics, statistically
significant periodicity estimates that are well sampled for the available
time-span, and single-exposure measurements with a relatively high
signal-to-noise ratio (e.g., S/N $\gtrsim 10$) in W1 or W2.
We expect to reliably classify close to one million periodic 
variable candidates.
The WVSDB will list derived properties such as periods, amplitudes, 
phased light curves, a vector of probabilities of belonging to
specific classes (see below) and from these, the ``most likely''
(or winning) class.

The classification probabilities will be {\it conditional class} 
likelihoods as defined
by equation~\ref{eq:prob}. By default, the RF classifier assigns the winning
class $C_j$ for an object with features $\mathbf{X}$ as that with
the highest probability $P(\mathbf{X}|C_j)$, with no margin for possible
classification error. For example, for the three broad classes
in our input training model, $P(\mathbf{X}|C_j)$ only needs to be $> 1/3$
to stand a chance of being assigned class $C_j$. Therefore, if the
probabilities for Algol, RR Lyrae, and W Uma are 
0.34, 0.33, and 0.33 respectively, the winning class is Algol. This  
assignment is obviously not significant in a relative sense
and we want to be more certain (or less ambiguous)
when reporting the most likely class.
Examining the conditional probability histograms 
in Figure~\ref{fig:rfprob}, a workable threshold for assigning a secure
classification (setting aside other biases; see below) may be $P > 0.6$.
The fractions of objects in our final validation
{\it test sample} (Section~\ref{cv}) initially classified as 
Algol, RR Lyrae, and W Uma that have $P > 0.6$ (and hence securely
classified) are $\approx83$\%, $\approx82$\%, and $\approx80$\% respectively. 
The remaining $\approx20$\% of objects with class probabilities 
$P\le0.6$ would be initially classified as ``unknown''.
This is a consequence of the ``fuzzy'' classification boundaries in our 
input training model.
Can these less probable (or more ambiguous) cases be classified into a
more secure (sub-)class in future?
Below we discuss an approach to mitigate this limitation.

\subsection{Mitigating Training Sample Biases and Limitations}\label{al}

It is known that RF classifiers trained using supervised methods
perform poorly outside their ``learned boundaries'', i.e., when extrapolating
beyond their trained feature space.
The RF training model constructed in Section~\ref{train} was tuned
to predict the classifications
of only three broad classes: Algol, RR Lyrae, and W Uma --
the most abundant types that could be reliably identified
given the WISE sensitivity and observing cadence.
Furthermore, this model is based on confirmed variables and classifications
from previous optical surveys (Section~\ref{tsamp}) which no doubt 
contain some incorrect labels, particularly since most of these studies
also used some automated classification scheme.
Therefore, our initial training model is likely to suffer from sample
selection bias whereby it will not fully represent all
the variable types that WISE can recover or discover
down to fainter flux levels and lower S/N ratios (Figure~\ref{fig:mags}).
Setting aside the three broad classes, our initial training model
will lead to biased (incorrect) predictions for other rare
types of variables that are close to or distant from the 
``fuzzy'' classification boundaries of the input model.

\begin{figure}[hbtp]
\begin{center}
\includegraphics[scale=0.6]{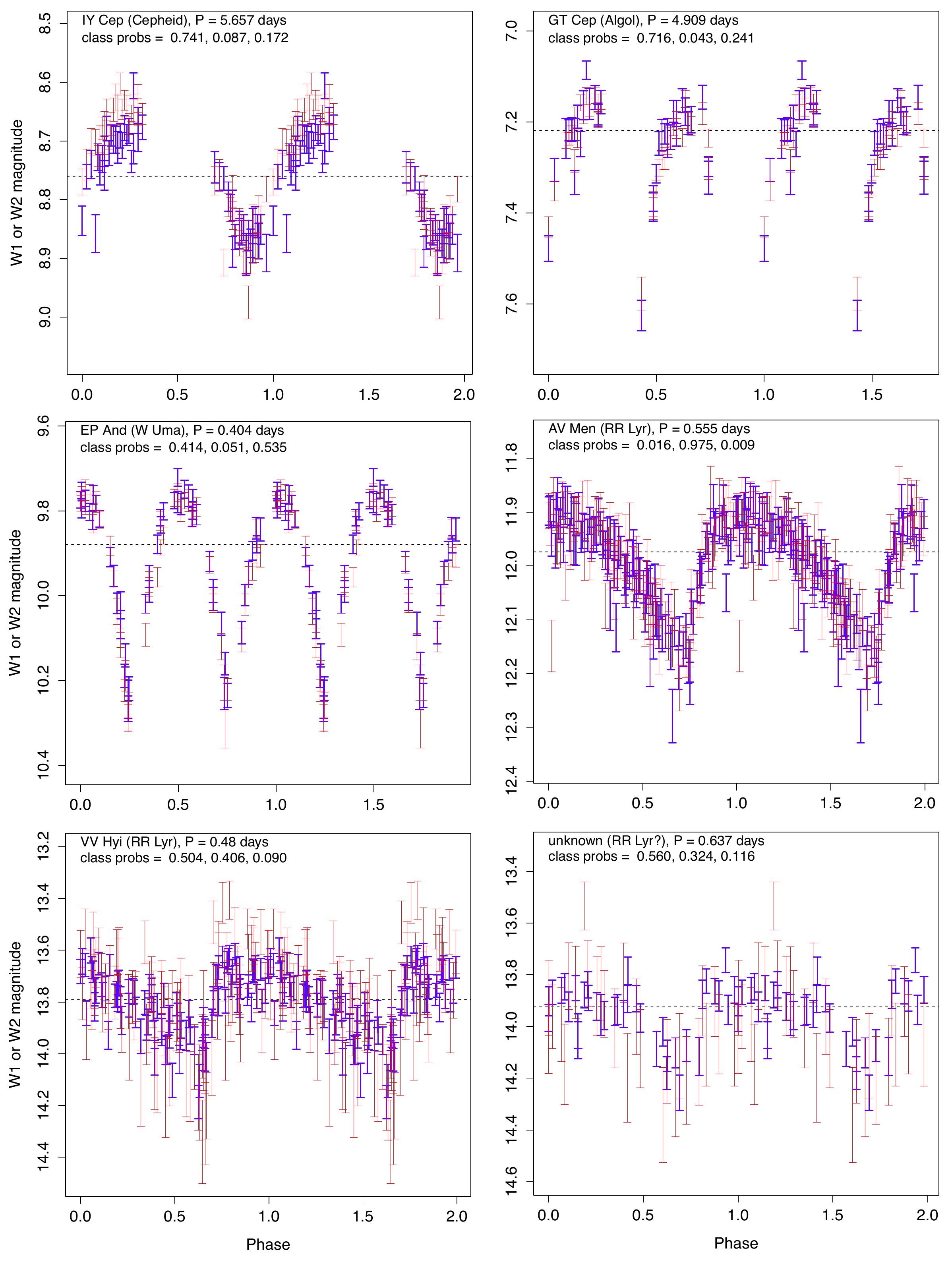}
\vspace*{-0.5cm}
\caption{Phased light-curves for five known variables and one
new object at bottom right (possibly a RR Lyrae) using WISE
single exposure photometry: W1 $=$ blue (thick bars);
W2 = red (thin bars). Horizontal dashed lines are median W1
magnitudes. Each panel shows the variable type (if known), period, and the
three class probabilities predicted by our initial RF training model: for
Algol, RR Lyr, and W Uma types respectively.}
\label{fig:eglcs}
\end{center}
\end{figure}

Figure~\ref{fig:eglcs} illustrates some of these challenges. 
Here we show example W1 and W2 light curves for a collection
of known variables from the literature (including one ``unknown'') 
and their predicted class probabilities using our input training model.
The known short-period Cepheid (top left) would have its period recovered
with good accuracy given the available number of WISE exposures that cover it.
However, it would be classified as an Algol with a relatively high probability. 
That's because our training sample did not include short-period Cepheids. 
Its period of $\sim5.6$ days is at the high end of
our fitted range (Figure~\ref{fig:period}) and overlaps with the Algol class.
For the given optimum observation timespan covered by WISE, the number of
short-period Cepheids after cross-matching was too low to warrant
including this class for reliable classification in future.
Better statistics at higher ecliptic latitudes (where observation timespans
are longer) are obviously needed.
The known Algol and one of the two known RR Lyrae in Figure~\ref{fig:eglcs}
are securely classified, although the known W Uma achieves a classification
probability of only 0.535 according to our training model. This would be tagged
as ``unknown'' if the probability threshold was 0.6 for instance.
The two lower S/N objects on the bottom row (a known RR Lyra and a fainter
variable we identify as a possible RR Lyra ``by eye'')
would also be classified as ``unknown'' according to
our initial model, even though their light-curves can be visually
identified as RR Lyrae. This implies that when the S/N is low and/or
the result from an automated classification scheme is ambiguous (following
any refinement of the training model; see below), visual inspection
can be extremely useful to aid the classification process.

%\placefigure{fig:eglcs}

We need a mechanism that fills in the undersampled regions of 
our training model but also improves classification accuracies for the
existing classes. \citet[][and references therein]{richards12}
presented methods to alleviate training-sample selection biases and we
use their concepts (albeit slightly modified)
to optimize and extend classifications
for the WVSDB. These methods fall under the general paradigm of
{\it semi-supervised} or {\it active learning}
whereby predictions and/or
contextual follow-up information for new data is used to
update (usually iteratively) a supervised learner to enable 
more accurate predictions.
\citet{richards12} were more concerned with the general problem 
of minimizing the bias and variance of classifiers
trained on one survey for use on predicting the outcomes for another.
Our training sample biases are not
expected to be as severe since our training model was 
constructed more-or-less
from the same distribution of objects with properties that we expect 
to classify in the long run. Our goal is simply to strengthen predictions
for the existing (abundant) classes as well as explore whether more
classes can be teased out as the statistics improve and more
of the feature space is mapped. Our classification process will
involve at least two-passes where the second pass (or subsequent passes)
will use {\it active learning} concepts to refine the training model.
A review of this classification process is given in Appendix~\ref{appB}.

\section{Conclusions and Future Work}\label{conc}

We have described a framework to classify {\it periodic} variable stars
identified using metrics derived from photometric
time-series data in the WISE single-exposure Source Databases.  
This framework will be used to construct an {\it all-sky} database
of variable stars in the mid-IR (the WVSDB), the first of its kind at 
these wavelengths. The reduced effects of dust-extinction will improve
our understanding of Milky Way tomography, the distribution 
of dark matter, stellar structure, and evolution in a range of
environments.

We identified several light-curve features to assist with the
automated classification of WISE periodic variables,
and found that Fourier decomposition techniques can be successfully
extended into the mid-IR to define features for unambiguously classifying
variable stars.
Guided by previous automated classification studies of variable stars, we
trained a machine-learned classifier based on the {\it random forest} method
to probabilistically classify objects in a seven-dimensional feature space.
Random forests satisfy our needs in terms of flexibility, ability to
capture complex patterns in the feature space, assessing feature importance,
their relative immunity
to outliers and redundant features, and for providing simple methodologies  
to support active-learning frameworks that can extend and refine
training models to give more accurate classifications. 

We constructed a training sample of 6620 periodic variables with 
classifications from previous optical variability surveys (MACHO, GCVS,
and ASAS) and found  that the most common types that separate rather well
and are reliably identified by WISE (given its sensitivity,
observing cadences and time-spans)
are Algols, RR Lyrae, and W Ursae Majoris type variables.
This sample was used to construct an initial RF training model
to assess classification performance in general and hence
whether our method was suitable for constructing a WVSDB.
From cross-validating a separate sample of
1653 pre-classified objects, our RF classifier achieves
classification efficiencies of 80.7\%, 82.7\%, and 84.5\% 
for Algols, RR Lyr, and W Uma types respectively, 
with 2-$\sigma$ uncertainties of $\sim \pm$2\%.
These are achieved at purity (or reliability) levels of $\gtrsim88\%$
where the only source of ``impurity'' to each specific class is
contamination from the other two contending classes. 
These estimates are similar to those of recent automated classification
studies of periodic variable stars in the optical
that also use RF classifiers.

Future work will consist of selecting good quality candidates for the
WVSDB, the computation of light-curve features, further selection
to retain periodic objects (above some statistical significance),
then construction of the WVSDB.
The three-class RF training model defined above will form the
basis for initially predicting the classes with associated probabilities
for the WVSDB. These probabilities will
be thresholded to secure the ``winning'' classes. This first 
classification pass will inevitably leave us with a large fraction 
of unclassified objects. Our input training model has considerable room
for expansion and improvement since during its 
construction, there were variable types that had to be removed
since they were either too scarce or
too close to an existing larger class to enable a reliable classification. 
Therefore, following the first classification pass,
we will refine the training model using an active-learning approach
(tied with contextual and/or follow-up information) where
the improved statistics will enable us to better map
and divide the feature space into more classes as well as sharpen the
boundaries of existing ones. This will appropriately handle the 
``known unknowns'', but WISE's all-sky coverage and sensitivity
offers a unique opportunity
to discover new and rare variable types, or new phenomena and sub-types in
existing classes of pulsational variables and eclipsing binaries.

\acknowledgments

This work was funded by NASA Astrophysics Data Analysis Program 
grant NNX13AF37G. We thank the anonymous referee for invaluable
comments that helped improve the quality of this manuscript. 
We are grateful to Max Kuhn for reviewing some of the details of our
analyses and descriptions of the algorithms implemented in
the \texttt{R} \texttt{caret} package.
This publication makes use of data products from the The Wide-field
Infrared Survey Explorer, which is a joint project of the University
of California, Los Angeles, and the Jet Propulsion
Laboratory/California Institute of Technology, funded by the National
Aeronautics and Space Administration.  Long-term archiving and access
to the WISE single-exposure database is funded by NEOWISE, which is a
project of the Jet Propulsion Laboratory/California Institute of
Technology, funded by the Planetary Science Division of the National
Aeronautics and Space Administration.  This research has made use of
the NASA/ IPAC Infrared Science Archive, which is operated by the Jet
Propulsion Laboratory, California Institute of Technology, under
contract with the National Aeronautics and Space Administration.


%% in parentheses, after the keyword, but they will not be verified.

{\it Facilities:} \facility{WISE}



\appendix

\section{Random Forest Training Algorithm}\label{appA}

Following the discussion in Section~\ref{train}, we outline
the steps used to train and tune the Random Forest
classifier as implemented in the \texttt{R} \texttt{caret} package.
To our knowledge, this is not documented elsewhere.
Concise generic descriptions of the algorithm exist in the literature,
but here we present the details for the interested reader.
These were inferred from our in-depth experimentation with
the software and dissecting the source code.

\begin{enumerate}
\item{Select a trial value for $m_{\rm try}$: the number of input features 
   to randomly select
   from the full set of $M$ features for determining the best split at each
   node of a tree.}
\item{Select an iteration $k$ in the 10-fold cross validation with a 
   training subsample consisting of $T=9N/10$ objects partitioned from the
   input training sample of $N$ objects.}
\item{Take a bootstrap sample from this training subsample of $T$ objects
   by randomly choosing $T$ times with replacement.} 
\item{Grow an un-pruned tree on this bootstrap sample where
   at each node of the tree, use the $m_{\rm try}$ randomly selected
   features to calculate the best split using the {\it Gini index} -- a measure 
   of the class inequality or impurity across a node. For some trial node $m$,
   this is defined as
   \begin{equation}
   G_m = \sum_{j\ne j^\prime}\left(\frac{N_j}{N_m}\right)
         \left(\frac{N_{j^\prime}}{N_m}\right),
   \end{equation}   
   where $N_m$ is the number of objects in node $m$ that are distributed
   amongst known classes $j=1,2,3...$ with respective numbers $N_j$.
   The best splitting hyperplane with respect to another adjacent node
   (that ultimately defines the new node $m$) is that which maximizes
   $G_m$.}
\item{Each tree is fully grown (up to its leaves) and not pruned.}
\item{For the given value of $m_{\rm try}$, repeat steps 3 to 5 for
   $N_{\rm tree}$ bootstrapped training samples
   to create $N_{\rm tree}$ trees in the {\it random forest}.}
\item{Predict classifications for every object in
   the $k$th ``hold-out'' subsample
   in the 10-fold cross-validation set using all the $N_{\rm tree}$ trees
   (see equation~\ref{eq:prob} in Section~\ref{train}).
   Compare these predictions to the known (true)
   classifications to compute the classification accuracy.  
   Store the average classification accuracy over all objects for the
   given $k$th iteration and trial value of $m_{\rm try}$.}
\item{Move to the next cross-validation iteration $k$ with new training
    subsample (step 2) and repeat steps 3 to 7.}
\item{When all 10 cross-validation iterations are done, average
    the classification accuracies
    from all 10 iterations. This represents the average classification accuracy
    for the given value of $m_{\rm try}$ selected in step 1.}
\item{Move to the next trial value of $m_{\rm try}$ and repeat steps 1 to 9.}
\item{When all trial values of $m_{\rm try}$ are tested,
    select the optimal value of $m_{\rm try}$ based on the highest
    average classification accuracy from all the cross-validation runs.}
\item{Using this optimal $m_{\rm try}$ value, construct the final
    RF model using {\it all} $N$ objects in the initial
    training sample from a final run of steps 3 to 6 with $T = N$.
    The final RF model consists of a ``\texttt{R} object'' that stores
    information for all the $N_{\rm tree}$ trees. This can then be used
    to predict the classifications for new objects (see Section~\ref{train}).}
\end{enumerate}

\section{Classification Plan and Active Learning Framework}\label{appB}

Below we given an overview of our classification plan for the WVSDB. This
uses two {\it active learning} methods to mitigate the limitations of our
initial training set discussed in Section~\ref{al}. These methods are
not new; they were formulated (with slight modifications) from the concepts
presented in \citet{richards12}. Details and results of overall performance
will be given in a future paper.

Depending on details of the manual follow-up of
``unclassifiable'' but good quality light curves, 
we expect at minimum, two passes in the classification process.
The first pass uses our initial RF training model to compute and store
the conditional-class
probabilities for each object (Algol, RR Lyrae, and W Uma).
In preparation for the {\it active learning} and
second classification pass, we compute the
averaged {\it RF proximity} metrics for
each object initially classified as ``unknown'' 
according to some probability cut (see Section~\ref{cpred}).
A proximity metric quantifies the relative separation between
any two feature vectors amongst the $N_{\rm tree}$ decision trees 
of the RF and is defined as
\begin{equation}\label{eq:prox}
\rho\left(\mathbf{X},\mathbf{X^\prime}\right) = \frac{1}{N_{\rm tree}}
      \sum_{i=1}^{N_{\rm tree}}I\left(T_i(\mathbf{X})=T_i(\mathbf{X^\prime})
      \right).
\end{equation}
This represents the fraction of trees for which two objects
with features $\mathbf{X}$ and $\mathbf{X^\prime}$ occupy the
same terminal node $T_i$ (leaf) where
$I()=1$ if the statement in parenthesis is true and
0 otherwise. We compute average proximity measures for {\it unknown} object  
$\mathbf{X}$ with respect to (i) all objects in the input training 
set $\{train\}$
and (ii) all other objects in the test set $\{test\}$ under study. These
are defined as 
\begin{equation}\label{eq:avprox1}
S\left(\mathbf{X}\right)_{train}=\frac{1}{N_{train}}
       \sum_{\mathbf{X^\prime}\in\{train\}}
       \rho\left(\mathbf{X},\mathbf{X^\prime}\right) 
\end{equation}
and
\begin{equation}\label{eq:avprox2}
S\left(\mathbf{X}\right)_{test}=\frac{1}{N_{test}}
       \sum_{\mathbf{X^\prime}\in\{test\}}
       \rho\left(\mathbf{X},\mathbf{X^\prime}\right)
\end{equation}
respectively.
The summation in equation~\ref{eq:avprox2} will be over a random subset of
objects in the test sample (both with known and unknown initial
classifications). This is to minimize runtime since we expect to
encounter at least a million variable (test) candidates. We expect to use
of order 20\% of the test objects.
We will be primarily interested in the ratio $R = S_{test}/S_{train}$.
$R$ will be larger for objects that reside in regions of
feature space where the test-sample density is higher relative to that in the
training set. Cuts in probability (from the first pass) versus $R$ and versus
$S_{test}$ can be used to identify regions occupied by 
new or missed classes whose statistics were scarce
when the training set was first constructed. Some analysis will be needed to
assign class labels to these ensembles of new objects. This can be done
by comparing to known variable light curves (again) from the literature  
(e.g., that could be associated with Cepheids, 
$\delta$ Scuti, $\beta$ Lyrae,    
or perhaps RR Lyrae subtypes), or, if insufficient information is available,
as possibly new interim classes to be labelled 
in future. The above is the first form
of active learning we will use to refine the training set.

Prior to the second classification
pass, we will also augment the input training set by adding the most
confidently classified test objects (Algol, RR Lyrae, and W Uma),
i.e., with a relatively high probability. This is a simple form of 
{\it self-training} (also known as {\it co-training}) in 
the machine learning toolbox. This will ``sharpen'' and possibly extend
classification boundaries
for the dominant classes and hence improve their prediction
accuracy. Following the addition of new classes (from the proximity analysis)
and new secure identifications to existing 
classes in the training set,
we will retrain the RF classifier on the new training data and reclassify
all objects in a second pass. Some fraction of objects will remain 
unclassified, but we expect their numbers to be considerably lower.
Nonetheless, these ``outliers'' will be potentially interesting objects
for further study.

As the RF classifier is refined according to the above scheme, we will also
explore improvements to the training model using {\it boosting} methods and
generalizations thereof such as {\it gradient boosted trees}
\citep[][and references therein]{buhlmann07, hastie09}.
Here, the training model is iteratively refit using re-weighted versions of
the training sample, i.e., progressively more weight (importance) is placed on
misclassified observations during training. These methods have been shown to
improve predictive performance over simple conventional RF classifiers.

\clearpage
%%% FIGS HERE

\begin{figure}[hbtp]
\begin{center}
\includegraphics[scale=0.8]{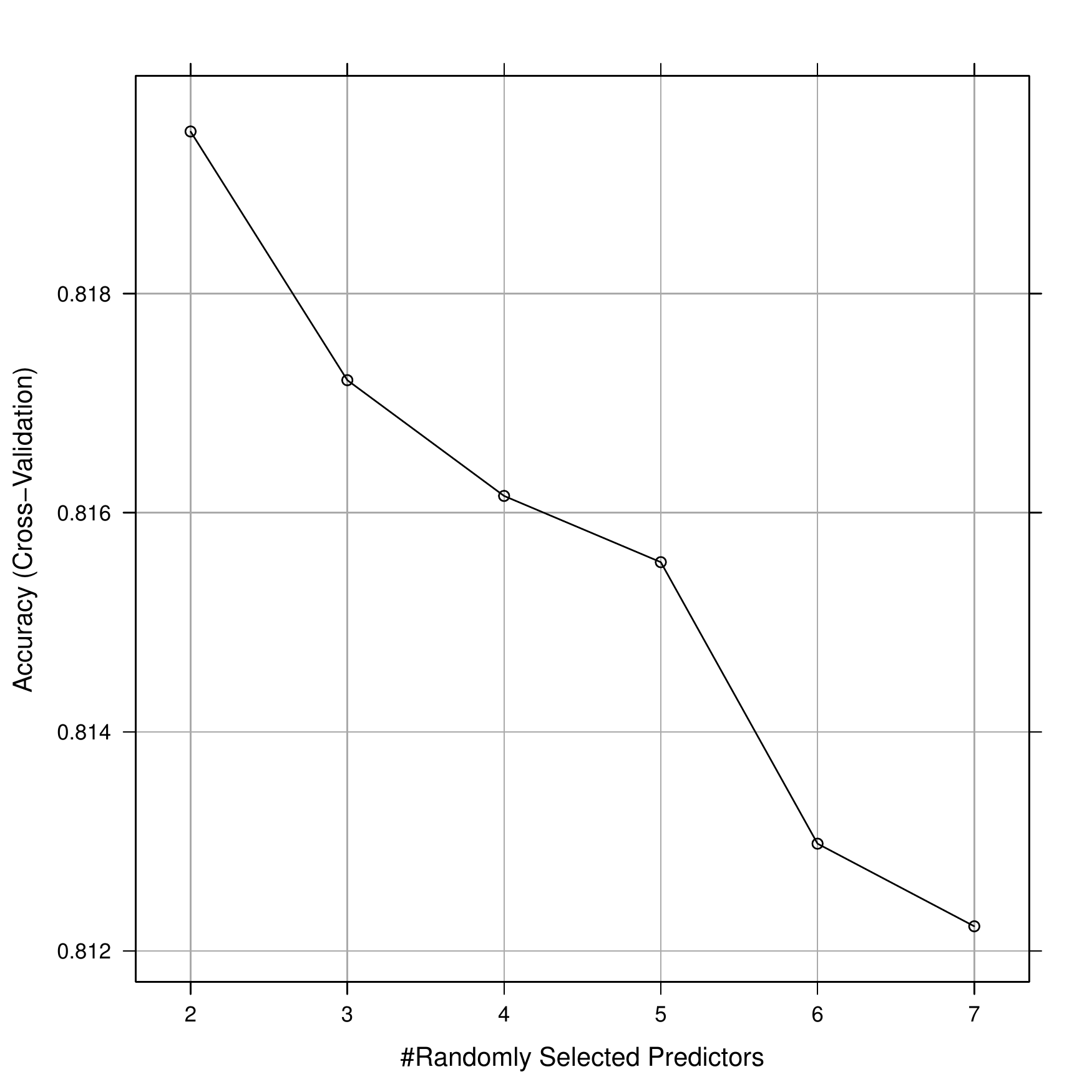}
\caption{Average classification accuracy from cross-validation as a
function of the number of randomly selected features to consider as
candidates for splitting at all nodes of a tree in the random
forest (i.e., the $m_{\rm try}$ parameter).}
\label{fig:mtry}
\end{center}
\end{figure}

%% Tables should be submitted one per page, so put a \clearpage before

\clearpage
%%% TABLES HERE

\begin{deluxetable}{lccccc}
\tabletypesize{\small}
\tablecaption{Classifier comparison\label{comp}}
\tablewidth{0pt}
\tablehead{
\colhead{Method} &
\colhead{Med. Accuracy}\tablenotemark{a} &
\colhead{Max. Accuracy}\tablenotemark{a} &
\colhead{Training time}\tablenotemark{b} &
\colhead{Pred. time}\tablenotemark{c} &
\colhead{{\it p}-value}\tablenotemark{d} \\
\colhead{} &
\colhead{} &
\colhead{} &
\colhead{(sec)} &
\colhead{(sec)} &
\colhead{(\%)}
}
\startdata
NNET  & 0.815  & 0.830  & 375.32  &  0.78  & 99.99   \\
$k$NN & 0.728  & 0.772  & 6.42    &  0.55  & $<0.01$ \\
RF    & 0.819  & 0.840  & 86.75   &  0.77  & \nodata \\
SVM   & 0.798  & 0.814  & 75.66   &  1.77  & 3.11    \\
\enddata
\tablenotetext{a}{Median and maximum achieved accuracies from a 10-fold
                  cross-validation on the training sample.}
\tablenotetext{b}{Average runtime to fit training model using parallel
                  processing on a 12-core 2.4 GHz/core Macintosh with 60 GB
                  of RAM.}
\tablenotetext{c}{Average runtime to predict classes and compute probabilities
                  for 1653 feature vectors in our final validation
                  {\it test sample} (Section~\ref{prep}).}
\tablenotetext{d}{Probability value for H0: difference in mean accuracy
                  relative to RF is zero.}
\end{deluxetable}

\end{document}